\documentclass[twocolumn]{aa}
\usepackage{natbib}
\usepackage{color}
\usepackage{ragged2e}
\usepackage[pdfpagelabels=false]{hyperref}	
\hypersetup{colorlinks=true,linkcolor=blue,citecolor=blue,filecolor=blue,urlcolor=blue,}
\usepackage[varg]{txfonts}
\usepackage{graphicx,rotating}
\usepackage[normalem]{ulem}
\usepackage{colortbl}
\usepackage{xcolor}
\usepackage{bbold}
\usepackage{stfloats}

\bibpunct{(}{)}{;}{a}{}{,} 

\definecolor{darkolivegreen}{rgb}{0.33, 0.42, 0.18}
\definecolor{salmon}{rgb}{0.95,0.5,0.25}

\begin{document}

\title{A new classification of ex-situ and in-situ Galactic globular clusters based on a method trained on Milky Way analogues in the TNG50 cosmological simulations}
\titlerunning{TNG50 Milky Way shake of globular clusters}
\author{Pierre Boldrini \inst{1}, Paola Di Matteo \inst{1}, Chervin Laporte \inst{1,4,5}, Oscar Agertz \inst{2}, Sergey Khoperskov \inst{3} and Giulia Pagnini \inst{6}}

\offprints{Pierre Boldrini, \email{pierre.boldrini@obspm.fr}}
\institute{$^{1}$ LIRA, Observatoire de Paris, Université PSL, Sorbonne Université, Université Paris Cité, CY Cergy Paris Université, CNRS, 92190 Meudon, France, \\$^{2}$ Lund Observatory, Division of Astrophysics, Department of
Physics, Lund University, Box 43, SE-221 00 Lund, Sweden, \\ $^{3}$ Leibniz-Institut für Astrophysik Potsdam (AIP), An der Sternwarte 16, 14482 Potsdam, Germany, \\
$^{4}$ Kavli IPMU (WPI), UTIAS, The University of Tokyo, Kashiwa, Chiba 277-8583, Japan, \\
$^{5}$ Institut de Ciències del Cosmos (ICCUB), Universitat de Barcelona, Martí i Franquès 1, E-08028 Barcelona, Spain, \\ $^{6}$ Université de Strasbourg, CNRS, Observatoire astronomique de Strasbourg, UMR 7550, F-67000 Strasbourg, France}
\authorrunning{Boldrini et al.}
\date{accepted by A$\&$A}

\abstract{We present a novel method combining existing cosmological simulations and orbital integration to study the hierarchical assembly of globular cluster (GC) populations in the Milky Way (MW). Our method models the growth and evolution of GC populations across various galactic environments as well as the dynamical friction and mass-loss experienced by these objects. This allows us to follow the trajectory of $\sim$18,000 GCs over cosmic time in 198 MW-like galaxies from TNG50. This cosmological-scale tracking of the dynamics of in-situ and ex-situ GC populations with such a large statistical sample allows us to confirm the presence of an overlap between the two populations in MW-like galaxies, occurring below an energy threshold of $E < -0.7 |E_{\rm circ}(r_{\rm hm}^{*})|$ where $E_{\rm circ}(r_{\rm hm}^{*})$ is the energy of a circular orbit at the galaxy’s stellar half-mass radius $r_{\rm hm}^{*}$. Our results challenge the validity of current classification schemes commonly adopted in the literature, which ultimately fail to provide a clear separation between the two populations. Instead, they tend to isolate only a subset of the ex-situ GCs. More precisely, we argue that it is highly unlikely to find in-situ clusters at $E > -0.7 |E_{\rm circ}(r_{\rm hm}^{*})|$, and that the real challenge lies in distinguishing the two populations below this energy threshold. In this context, we provide new predictions regarding the origins of the MW’s GCs observed with Gaia, as well as a comparison with existing literature. Additionally, we highlight that even if ex-situ clusters share a common origin, they inevitably lose their dynamical coherence in the $E$-$L_{z}$ space within MW-like galaxies. We observe a dispersion of GC groups as a function of $E$ and $L_{z}$, primarily driven by the evolution of the galactic potential over time and by dynamical friction, respectively.}

\maketitle



\section{Introduction}

The study of the Milky Way (MW) provides a unique perspective on galaxy assembly history, serving as a crucial benchmark for theories of galaxy formation and evolution. Among the various stellar systems within the MW, globular clusters (GCs) play a fundamental role as ancient relics that encode essential information about the galaxy’s assembly history. A key aspect of MW GCs is their diverse origins: while some are thought to have formed in-situ within the progenitor of the MW, others were accreted (ex-situ) through the mergers of satellite galaxies \citep{Forbes10,Massari19,Kruijssen19,Helmi20,vandenBergh2012,Leaman2013,Mackey2005}. Studies of GCs in external galaxies, such as the analysis of accreted GCs in M31 by \cite{Andersson2019}, provide complementary insights into the processes of GC accretion and the signatures of past merger events.

Identifying and characterizing these two populations is essential for reconstructing the accretion history and evolutionary pathway of our Galaxy. This has become possible thanks to the full six-dimensional phase-space information available for nearly all Galactic GCs from the third data release of the Gaia mission \citep{Gaia21}, which has provided unprecedented insights into their dynamics \citep{Vasiliev21}. In principle, these data contain fundamental information about the origin of the clusters themselves, that is whether a given GC formed within the MW or originated in a satellite galaxy that was later accreted. However, the question that the community is currently facing is how to analyse and interpret these data in order to discern the origin of Galactic GCs. Various methods \citep[see, for example,][]{Massari19, Forbes10, Kruijssen20, Kruijssen19E, Malhan22, Myeong18} have been developed to classify GCs as in$-$situ or ex$-$situ, leveraging their kinematic, spatial, age and chemical abundance properties with the latter being provided in part by spectroscopic surveys \citep[e.g. APOGEE][]{Majewski2017}. In particular, the distribution of GCs in integral-of-motion spaces, such as the energy-angular momentum $E$-$L_{z}$ plane, is in-homogeneous. This in-homogeneity suggests the presence of groups of GCs with distinct spatial and kinematic properties. Clusters located close to each other in this integral-of-motion space are thought to share a common origin, potentially reflecting different formation histories, either in-situ or through accretion \citep{Chen24, Belokurov23, Massari19, Malhan22, Pfeffer20, Sun23}. More recently, novel approaches which have complemented analysis in the $E$-$L_{z}$ plane with mean chemical abundance ratios such as [Al/Fe] have been also introduced to refine the quest of accreted and in-situ GCs \citep{Belokurov24}.

However, \cite{Pagnini23} challenged the idea of a dynamical coherence among accreted GCs, emphasizing that even the debris from a single merger event can span a wide range in the $E$-$L_{z}$ space, as well as in all kinematic spaces used so far, including actions space. They argue that the assumption that accreted clusters should exhibit a clear dynamical coherence, that is a tendency to cluster in kinematic spaces, remains unproven, and that while physically motivated by the conservation of energy and angular momentum in static, axisymmetric potentials, this expectation may not hold in more realistic, time-dependent galactic environments. Indeed, $E$ and $L_{z}$ are not conserved quantities in a time-evolving non-axisymmetric galaxy \citep{2008gady.book.....B}, due to the significant growth of the galaxy’s mass over time and the mergers of massive satellites, which perturb both the accreted and in-situ components of the galaxy. As a result, relying solely on the present-day ($z = 0$) distribution in the $E$-$L_{z}$ space to infer the origin of GCs can be misleading, requiring additional information and potentially the study of their temporal evolution. The results in \cite{Pagnini23} are consistent with findings for field stars, for which it is also challenging to uncover their nature using kinematic spaces \citep{2017A&A...604A.106J,2022ApJ...937...12A}, even when supplemented with abundance information \citep{2023A&A...677A..89K,2023A&A...677A..90K,2024A&A...690A.136M}. The difficulty encountered in establishing the origin of the stars and GCs in our Galaxy by making use of energy and angular momentum (eventually supplemented by other properties) is illustrated for example by the case of Omega Centauri (NGC 5139), which is likely the former nuclear star cluster of an accreted galaxy, as suggested - among other unusual properties - by its broad metallicity spread \citep{1999Natur.402...55L,2000LIACo..35..619M,2000A&A...357..977C,2003MNRAS.346L..11B,2004MNRAS.350.1141T,2003ApJ...589L..29T}. While some classifications  might place it among in-situ GCs \citep{Belokurov24, Chen24}, others support an ex-situ origin for this system \citep{Massari19, Forbes10, Pfeffer21}.

Studying the dynamics of GCs over cosmological timescales and within cosmological environments is crucial for understanding the formation and evolutionary history of galaxies like the MW. However, due to the vast spatial and temporal scales involved, simulating these objects presents significant challenges, leading to the adoption of various approaches, each with their inherent limitations (for reviews, see \cite{Beasley20, Renaud20}). Three main strategies have emerged for modeling GC dynamics, incorporating increasingly sophisticated models of GC formation. A common approach relies on idealized galaxy simulations with pc resolution, which enables the study of clustered star formation at high redshift. However, most of these simulations do not robustly reproduce GCs in terms of their sizes, masses, or chemical properties \citep{Lahen20, Li22, Deng24, Andersson24}. Besides, these simulations can only model faint galaxies and in small numbers (typically one or two) and are limited in their evolutionary timescale (typically up to 1 Gyr) due to their high computational cost. Another strategy involves high-resolution cosmological zoom-in simulations, which enhance spatial resolution and allow for the study of GC formation in a more realistic cosmological setting \citep{Kravtsov05, Li17, Kim18, Ma20, Meng22, Sameie23, Dubois21}. While these simulations achieve extremely high mass resolution (resolving dwarf galaxies) and spatial resolution (down to 10 pc), they are typically restricted to high redshifts ($z = 3-5$). To overcome these spatial and temporal limitations, one approach involves incorporating semi-analytical models within hydrodynamical simulations to form and track the evolution of GCs \citep{Pfeffer18, ReinaCampos22, Grudic23, Newton24, Rodriguez23}. However, this does not reduce the computational cost of hydrodynamical simulations, which remain limited to a small number of galaxies (typically 1$-$21 MW-like galaxies). To achieve larger statistical samples, an alternative approach is to apply "tagging" techniques in post-processing \citep{Bullock05,Cooper10, Laporte13,Renaud17,Ramos-Almendares20, Halbesma20, Park22, Doppel23, Chen23, Creasey19}. In these methods, GCs are "painted" onto simulation particles (either dark matter (DM) or stars) following specific prescriptions. These techniques are computationally inexpensive, as they do not require rerunning simulations to modify the GC formation and evolution model parameters, and they enable the study of a large number of galaxies (up to $N=8000$) in diverse environments. Although various tagging techniques exist to study GCs, either by tagging GCs at multiple epochs to analyze their properties within galaxies, or by inferring their dynamics from star or DM particles within simulations, no study has applied these methods on a large statistical sample of MW–like galaxies in a self-consistent manner. Our approach stands out by applying it to nearly 200 MW–like galaxies from TNG50, enabling a statistically robust analysis of GC dynamics from formation to $z=0$. Therefore, developing a complete picture of GC dynamics requires combining tagging techniques in cosmological hydrodynamical simulations with orbit integration methods. In principle, coupling these approaches could allow for the self-consistent tracking of GC dynamics within a cosmological environment over a Hubble time.

In this paper, we revisit the challenge of distinguishing between in-situ and ex-situ GC populations in MW-like galaxies, focusing on their identification within the $E$-$L_{z}$ space. Our analysis is conducted in a full cosmological context, including satellite galaxy accretions and a time-evolving potential for the host galaxy, both of which affect GC dynamics. To achieve this, we follow the spatial evolution of a population of tens of thousands of GCs within a cosmological framework, leveraging orbital integration techniques in combination with the TNG50 simulation. The paper is structured as follows: Section 2 outlines our methodology, which describes our post-processing orbital integration method for GCs in cosmological simulations. In Section 3, we present and discuss our results, comparing them with Gaia data for the MW’s GC system. Finally, Section 4 provides our conclusions.

\section{Methods: combining orbital integration methods in post-processing with TNG50 simulation}

To track the dynamics of GCs over cosmological timescales in a galaxy like the MW, it is essential to consider its merger history, which plays a key role in constructing the observed GC population at $z=0$. By providing access to the evolutionary history of 198 MW analogues, the hydrodynamical cosmological simulation TNG50 \citep{Nelson19b,Nelson19a,Pillepich19} enables us to reconstruct the time-evolving gravitational potentials that account for both the evolution of the MW and its environment through satellite galaxy accretion. TNG50\footnote{Available at \url{https://www.tng-project.org/}} is a high-resolution cosmological hydrodynamical simulation with a 51.7 Mpc box and $2\times2160^3$ particles, reaching a baryonic mass resolution of $8.4\times10^4$ M$_{\odot}$ and $4.5\times 10^5$ M$_{\odot}$ for DM. With a gravitational softening of 288 pc at z=0, it includes detailed models for star formation, chemical enrichment, black hole growth, AGN feedback, and galactic winds \citep{2017MNRAS.465.3291W,2018MNRAS.473.4077P}. Besides, \cite{Pillepich24} provide masses and half-mass radii at different redshifts of these MW-analogues and their associated satellites. By assigning a GC population to each MW progenitor and its satellite population at high-z, we can follow their dynamical evolution through orbital integration using the publicly available code \texttt{galpy}\footnote{Available at \url{https://github.com/jobovy/galpy}} \citep{Bovy15}, taking into account both dynamical friction and mass loss for these time evolving MW-like potentials from $z=3$ to the present-day. In the following, we first describe the MW-like potentials, then how we defined initial conditions for our GCs systems at high-z, and finally how we performed the orbit integration.

\begin{figure}[!t]
\centering
\includegraphics[width=\hsize]{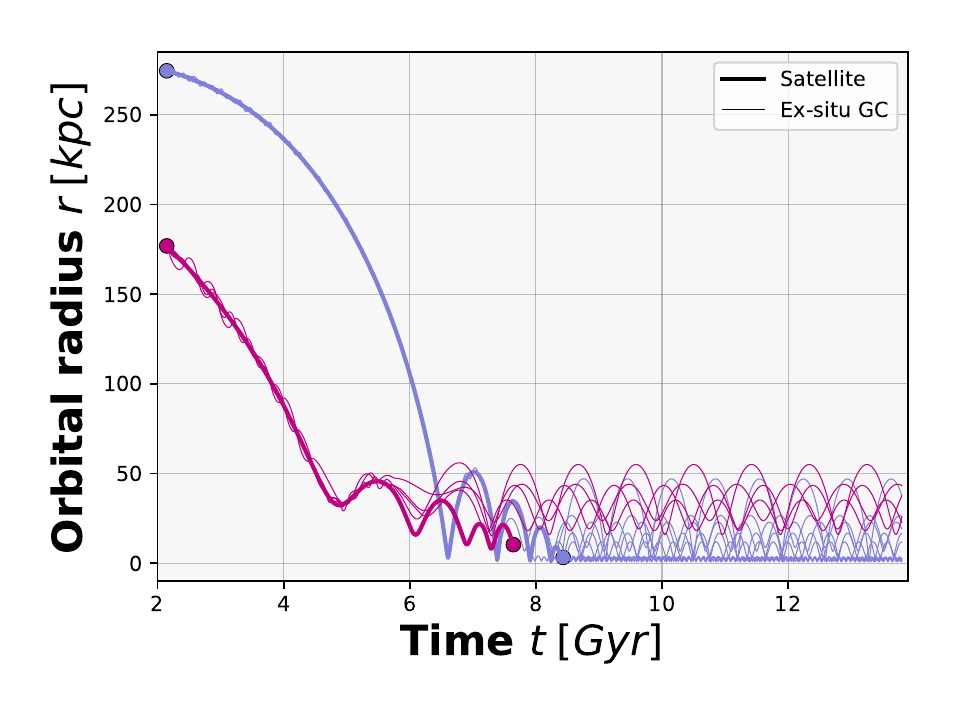}
\caption{Orbital radius as function of time for two merging galaxies (IDs 383325 and 299509) with their associated GCs. The points mark the formation and merger times of the satellites from TNG50. During the initial phase, ex-situ GCs orbit within the reference frame of their satellite galaxy until their orbital energy becomes positive ($E > 0$). Once this condition is met, GCs are transferred to the MW's reference frame while maintaining the moving potential of the satellite galaxy if it has not yet merged.}
\label{fig2}
\end{figure}

\subsection{TNG50 Milky Way-like potentials}
\label{section21}

\begin{figure}[!t]
\centering
\includegraphics[width=\hsize]{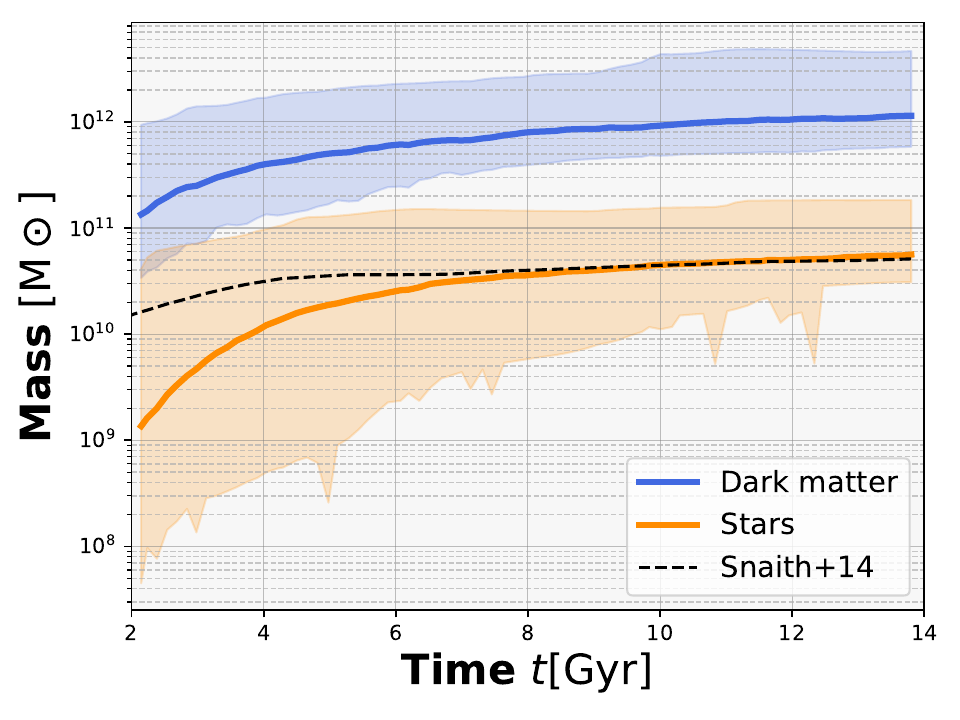}
\caption{The mass evolution of the stellar and DM components of the MW for our full sample of 198 MW analogues from TNG50 is shown. Up to 7 Gyr backwards in time, the mean stellar mass of the TNG50 MW analogues is in good agreement with the mass evolution of our Galaxy derived by \cite{Snaith2015}.}
\label{figA3}
\end{figure}

The 198 MW analogues identified in the TNG50 simulation have global properties, such as stellar mass between $10^{10.5}$ $\rm{M_{\odot}}$ and $10^{11.2}$ $\rm{M_{\odot}}$ and large-scale environment, that is similar to those of our Galaxy. Up to 7 Gyr backwards in time, the mean stellar mass of the TNG50 MW analogues is in good agreement with the mass evolution of our Galaxy derived by \cite{Snaith2015} (see Figure~\ref{figA3}). We extract the structural parameters (masses and half-mass radii) of the MWs and their associated merging satellites across 75 snapshots from $z=3$ to 0. Actually for TNG50, snapshots at all 100 available redshifts, galaxy catalogs at each snapshot and merger trees are released. We focus on identifying all the satellites that merged with our MW-like galaxies during this period, retaining only those with a stellar mass greater than $10^7$ $\rm{M_{\odot}}$, in line with the TNG50 mass resolution ($4.5 \times 10^5$ $\rm{M_{\odot}}$ for DM particles and $8.5 \times 10^4$ $\rm{M_{\odot}}$ for the stellar component). We modeled the  gravitational potentials of MW–like galaxies using a Hernquist profile \citep{Hernquist90} for the stellar component and an NFW profile \citep{Navarro97} for the DM halo. The parameters were derived from the properties of galaxies in the TNG50 simulation, which provides both total masses and half-mass radii for the stellar and DM components. We then computed the corresponding scale radii of the Hernquist and NFW profiles from the given half-mass radii. These MW-like potentials evolve over time, and the evolution of the masses and scale radii is incorporated into our model. Specifically, we update the potential 75 times during the orbit integrations. Satellite galaxies are also modeled with the same potential components.

To summarize, from the masses and half-mass radii provided by TNG50, we can derive $198\times 75$ MW-analogues composite stellar+DM gravitational potentials. Each of these MW analogues is surrounded by its processions of satellite galaxies (on average 4.94 merging satellites/MW-like progenitor are found at $z=3$), which are also modeled by composite stellar+DM potentials.

\subsection{Initial conditions for globular clusters}

Once the gravitational potentials of the MW progenitor galaxies and their satellites are defined at high redshift, we describe how a GC system was assigned to each of these galaxies. The initial number of GCs in galaxies is defined by the relation from \cite{Burkert20}, which correlates $N_{\text{GC}}$ with the virial mass of the halo:
\begin{equation}
\langle \log N_{\text{GC}} \rangle = -9.58 \pm 1.58 + (0.99 \pm 0.13) \log \frac{M_{\text{vir}}}{M_{\odot}}.
\end{equation}

\begin{figure*}[!t]
\centering
\includegraphics[width=\hsize]{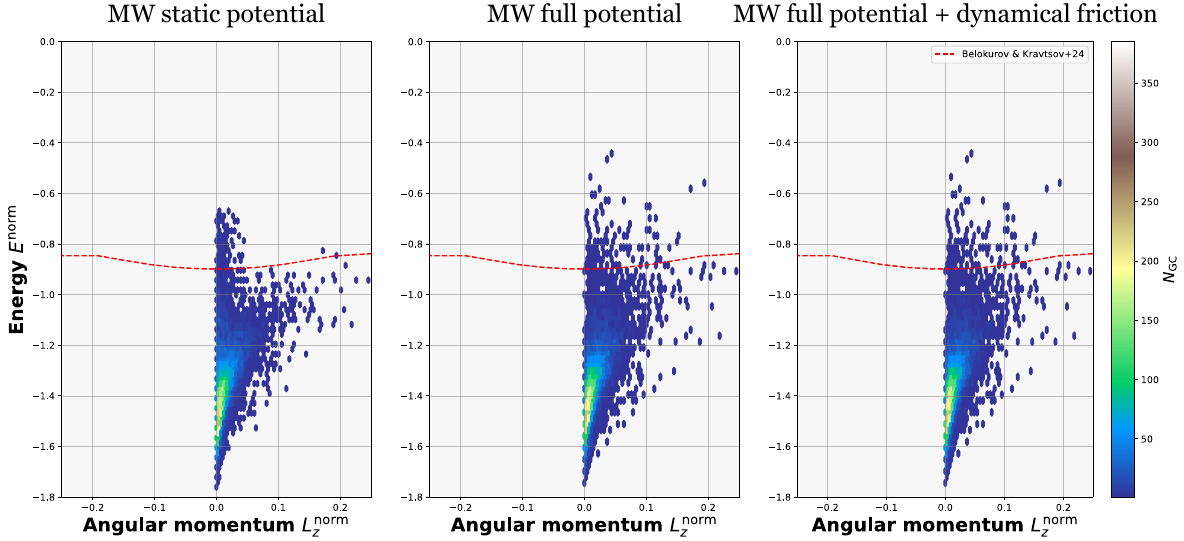}
\caption{In-situ populations at $z=0$: Normalized total energy as function of the normalized z-component of the angular momentum at $z=0$ for all our in-situ GCs for the 198 MWs with three different descriptions for the MW environment (see section 2.1 for the definitions of potentials). The hexagonal bins represent at least 1 GC. The red dashed line represents the \citep{Belokurov24} limit.}
\label{fig3}
\end{figure*}

\begin{figure*}[!t]
\centering
\includegraphics[width=\hsize]{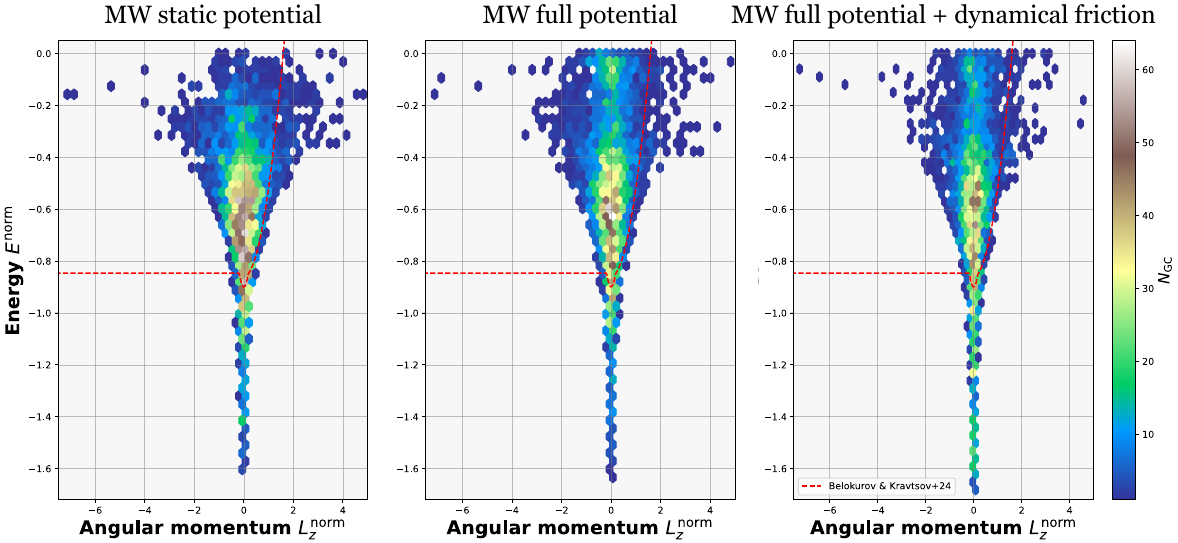}
\caption{Ex-situ populations at $z=0$: As in Fig~\ref{fig3} but for our ex-situ GCs for the 198 MWs.}
\label{fig4}
\end{figure*}

\begin{figure}[!t]
\centering
\includegraphics[width=\hsize]{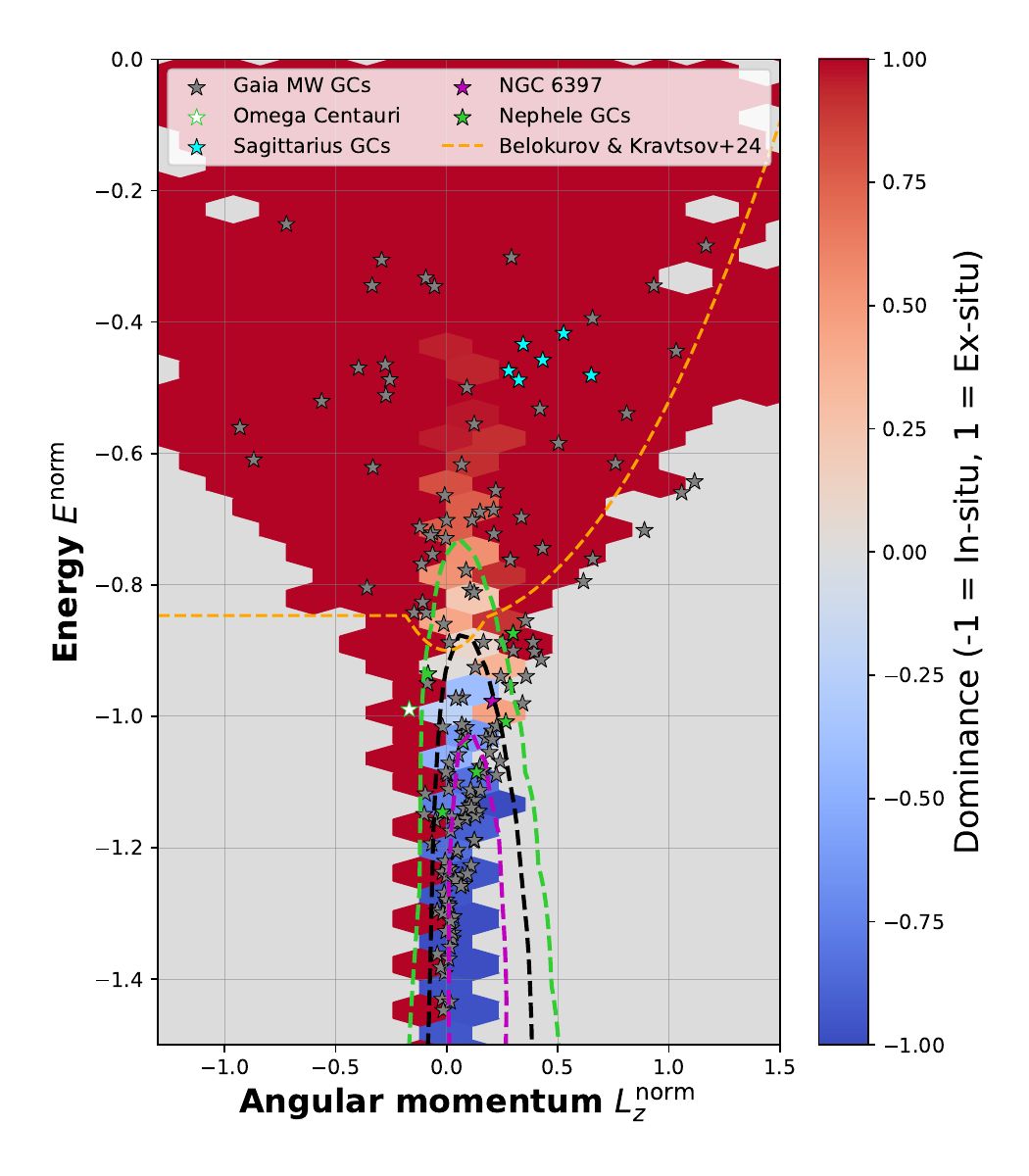}
\caption{Comparison with Gaia MW GCs: Normalized total energy as function of the normalized z-component of the angular momentum at $z=0$ for both in-situ and ex-situ GCs for the 198 MWs in the full MW potential + dynamical friction. The decision boundaries in dashed lines corresponding to probabilities of 0.05 (magenta), 0.5 (black), and 0.95 (green) are plotted to illustrate the separation between the two populations. Nephele GCs are the six GCs brought by the progenitor of Omega Centauri identified by \cite{Pagnini25}. The 0.5 boundary results in an error of 1.35$\%$ for in-situ GCs and 11.2$\%$ for ex-situ GCs. The extreme boundary at 0.95 (0.05) excludes 99.72$\%$ (97.3$\%$) of in-situ (ex-situ) GCs while identifying 81.2$\%$ (76.4$\%$) of ex-situ (in-situ) GCs. We establish that between these two limits, distinguishing the two populations is not feasible with previous methods due to the presence of a mixing zone. The orange dashed line represents the McMillan version of \citep{Belokurov24} limit.}
\label{fig5}
\end{figure}

\begin{figure*}
\centering
\includegraphics[width=\hsize]{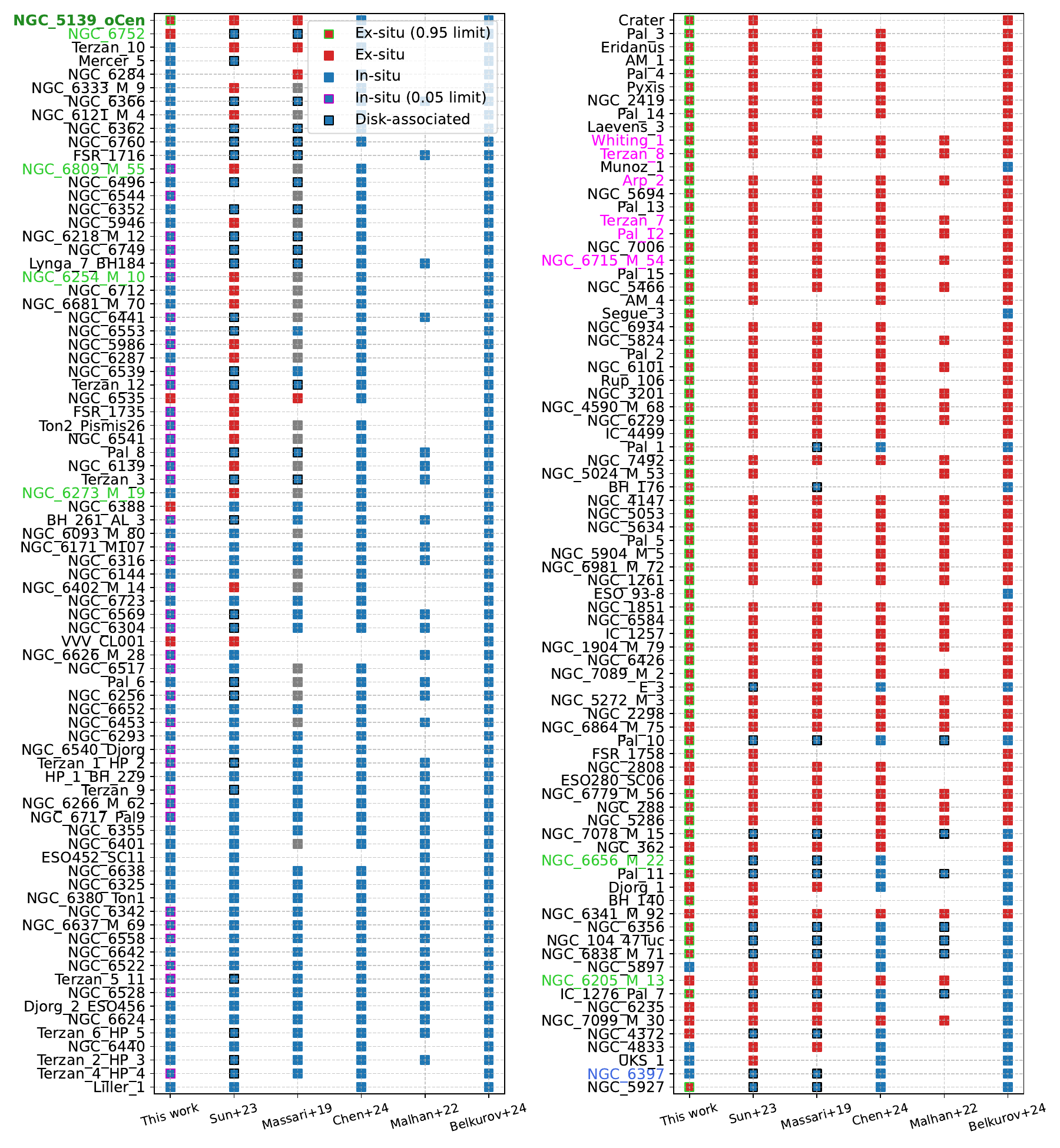}
\caption{Origin of MW GCs: Each point represents a GC and is color-coded according to its origin. GCs are sorted from the most to the least bound in energy. In-situ clusters (blue) with black edges are those associated with the MW disk by other studies. We also highlight the six GCs still associated with the Sagittarius dwarf galaxy (magenta), NGC 6397 (blue), and Omega Centauri (bold green), along with its six candidate clusters identified by \cite{Pagnini25} (light green). \cite{Massari19} GCs whose origin remains unidentified but which have been classified as 'low-energy' GCs are in grey.}
\label{fig5a}
\end{figure*}

\begin{figure}[!b]
\centering
\includegraphics[width=\hsize]{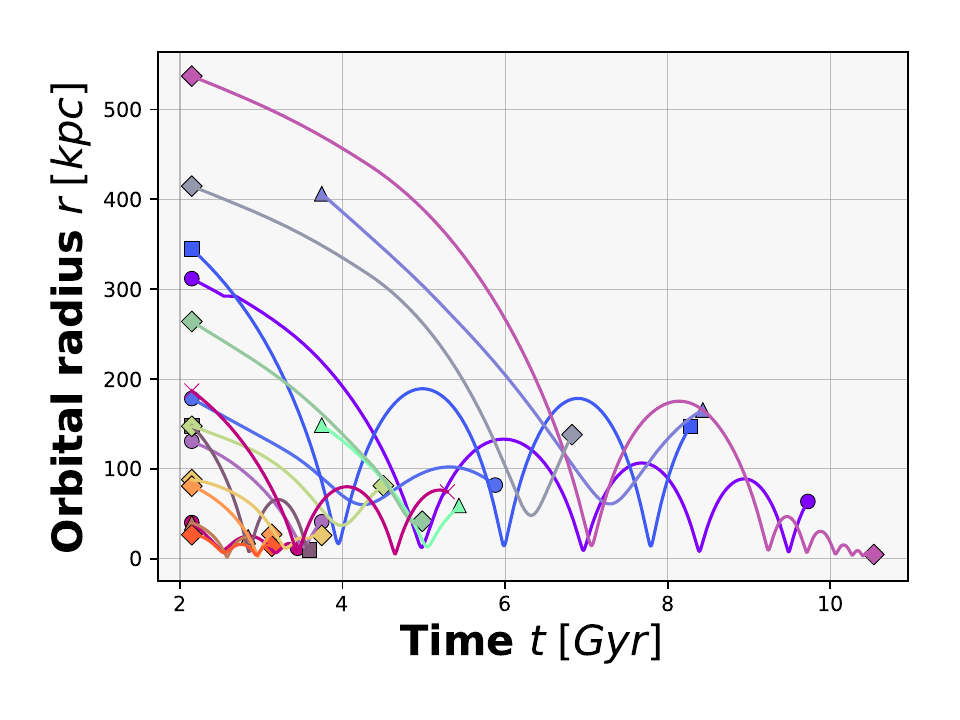}
\caption{Orbital evolution of merging satellites from five MW galaxies (IDs 342447, 372754, 372755, 388544 and 392277) in our sample, described by a full potential + dynamical friction. The points mark the formation and merger times of the satellites in the TNG50 simulation.}
\label{fig6}
\end{figure}

\begin{figure*}[!t]
\centering
\includegraphics[width=\hsize]{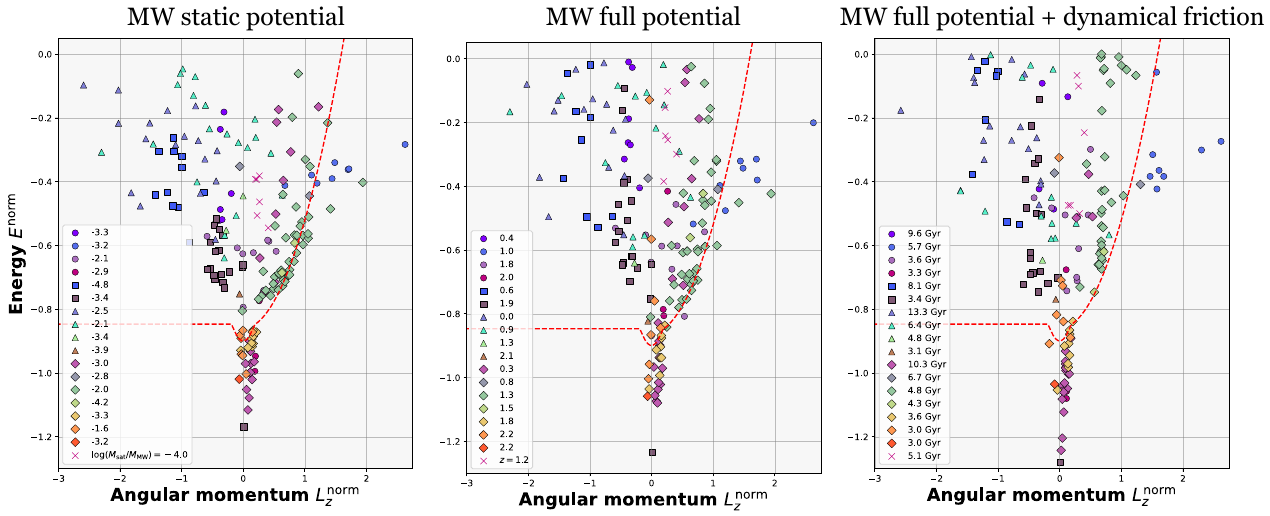}
\caption{E - $L_{z}$ decoherence of ex-situ GC groups due to evolving MW potentials and dynamical friction: Normalized total energy as function of the normalized z-component of the angular momentum at $z=0$, colour-coded according to their associations with satellites from the same five MW galaxies shown in Figure~\ref{fig6}. The captions provide information on the mass ratio between the satellite galaxy and the MW at the time of the merger (left panel), the redshift (middle panel), and the merger time (right panel).}
\label{fig7}
\end{figure*}

Although this relation does not explicitly include the star formation rate or GC formation efficiency, it likely reflects their cumulative effects as functions of halo mass. This relation depends on the DM mass of the galaxies at the redshift where the clusters are tagged. We included uncertainties as random noise in our modelling. We note that applying this relation, which is calibrated on nearby galaxies, to galaxies at z=2–3 introduces some uncertainty. Indeed, \cite{2019MNRAS.488.5409C} show that this relation can evolve by up to a factor of 10 from $z=3$ to $z=0$. However, such models depend on assumptions about GC formation physics and merger histories that remain unconstrained by direct observations at high redshift. Given the lack of observational data on GC populations at $z>1$, we adopt the \citet{Burkert20} relation as a pragmatic, observation-driven approximation. We generate a GC distribution at $z=2$ in the progenitors of the MW for the in-situ population, and at $z=3$ in the progenitors of merging galaxies for the ex-situ population. To estimate the number of in-situ GCs in MW–like galaxies, we chose to tag them at $z=2$, an epoch when the host galaxy’s mass approaches its present-day value. Indeed, the stellar halo was likely accreted between 9 and 11 Gyr ago \citep{2019A&A...632A...4D,2020MNRAS.492.3631M}, and cosmological simulations predict that the MW’s mass reached its asymptotic value around 9–10 Gyr ago \citep{2007ApJ...657..262D,2010MNRAS.406.2312L,Diemer13}. The choice of $z=2$ to assign GC systems is observationally motivated to reproduce the observed stellar mass–GC number relation at $z=0$, based on two major arguments. First, the GC–halo mass relation becomes valid for a larger number of galaxies at this epoch. This relation is particularly reliable once the galaxy’s halo mass exceeds a certain threshold ($M_{\rm DM}>5 \times 10^9$ $\rm{M_{\odot}}$). At $z=3$, many MW–type galaxies have only just reached this mass threshold. They are still in a rapid growth phase through mergers and accretion. By $z=2$, most of these galaxies have already accumulated a mass close to their final value. This means the GC–halo mass relation starts to apply more systematically at $z=2$, whereas it is more noisy at $z=3$. This leads to an underestimation of the total number of in-situ GCs at this higher redshift. Second, in a parallel simulation where clusters are tagged earlier, at $z=3$, we observe that not only is the initial number of clusters half as large, but their destruction rate by $z=0$ is three times higher. This is explained by the longer evolution time in a growing potential, where tidal effects promote disruption. This biases not only their initial number but also the final observable number at $z=0$. As a result, the dominance map in energy–angular momentum space is strongly biased in the $z=3$ case, artificially overestimating the fraction of ex-situ clusters due to the lack of surviving in-situ clusters. Table~\ref{tab1} shows that this model, which recovers only 23.5\% of simulated in-situ clusters, represents the worst-case scenario. Finally, tagging in-situ GCs at $z=2$ therefore ensures a more realistic estimate of the total number of in-situ clusters observable at $z=0$. In cases where our satellites were not yet formed at $z=3$ in the simulation, we tag the GCs at the highest possible redshift.

Next, in each sufficiently massive galaxy ($M_{\rm DM}>5 \times 10^9$ $\rm{M_{\odot}}$) that at least one GC is present, we derive the initial positions and velocities of in-situ GCs from stellar distribution functions based on our sample of progenitor galaxies and ex-situ GCs from those of the TNG50 stars. For in-situ populations, we use the publicly available code \texttt{AGAMA}\footnote{Available at \url{https://github.com/GalacticDynamics-Oxford/Agama}} to construct equilibrium galaxy models composed of a stellar bulge and a DM halo for our $z=2$ MW progenitor \citep{AGAMA}. For each system, a self-consistent potential is computed via iterative modelling, from which a phase-space distribution of stellar particles is sampled. We select a subset of star particles located within the stellar half-mass radius, forming an initially spherical distribution, to represent in-situ GCs. This choice is motivated by the assumption that in-situ GCs form preferentially in the central regions of galaxies, where the gas density and thus the likelihood of massive star cluster formation is highest. Within this spherical region, we then select only those particles with angular momentum $L_z > 0.6 \; L_{\rm circ}$, where $L_{\rm circ}$ is the angular momentum of a circular orbit at the same energy.This kinematic selection impose a disc-like initial configuration for the in-situ population. Indeed, the stellar sample at $z=2$ in the TNG50 MW analogues does not contain enough star particles with $L_z > 0.6 \; L_{\rm circ}$ within the stellar half-mass radius to match the number of in-situ GCs expected from the empirical relation of \cite{Burkert20}. For ex-situ GCs, we only randomly select $N_{\text{GC}}$ stars within the stellar half-mass radius of the galaxies. Therefore, our accreted GCs inherit the positions and velocities of the stars present in TNG50 satellites. 

We assign a mass of $10^6$ $\rm{M_{\odot}}$ and a half-mass radius of 10 pc to all GCs. This arbitrary choice is motivated by the fact that it allows us to recover the typical mass ($\sim 10^5$ $\rm{M_{\odot}}$) and size values of MW GCs today rather than modeling the full mass spectrum of MW clusters. In fact, a mass of $6 \times 10^6$ M$_{\odot}$ corresponds to the upper limit of the stellar mass of Galactic GCs 12 Gyr ago \citep{Baumgardt19}. In Table~\ref{tab1}, we explored the impact of varying the mass choice on our results, which only affects the calculation of dynamical friction and mass loss. It was found that our results remain unchanged within a reasonable initial mass range for GCs ($0.5 - 5 \times 10^6$ $\rm{M_{\odot}}$). Increasing or decreasing the mass affects only the in-situ population, decreasing or increasing its GC number, respectively (see Table~\ref{tab1}). However, we did not explore even lower masses for the following reasons. Our objective is to tag progenitors of present-day MW GCs. These must be sufficiently massive at formation to survive tidal disruption over a Hubble time. Lowering the initial GC mass below $10^5\,M_\odot$ would lead to enhanced disruption, especially in the central regions of galaxies, and very few GCs would survive until $z = 0$. In our lowest-mass test ($5\times10^5$ M$_{\odot}$), the in-situ GC population becomes significantly harder to identify using our SVM-based classification. The depletion of the central regions due to increased disruption reduces the number of surviving in-situ clusters. Consequently, the classification accuracy drops to 41$\%$, compared to 76.4$\%$ in our fiducial model. For these reasons, and because our goal is to model the surviving MW GC population at $z=0$, we chose a fiducial mass of $10^6$ M$_{\odot}$. Moreover, dynamical friction for in-situ GCs is inefficient over our timescale (11 Gyr) due to the large mass ratio between the MW's enclosed mass and the GC masses. While the GC half-mass radius is kept constant during our orbit integrations, we account for mass loss over time \cite{Kruijssen11}. For this mass loss model, we adopted the following parameters: $\gamma = 0.62$ and $t_{0,\odot} = 21.3$ Myr, which correspond to clusters with masses around $\sim 10^6$ M$_{\odot}$. Therefore, with these parameters, we aim to describe an average behavior over time. We also performed orbital integrations varying the parameters of this mass loss model. Table~\ref{tab1} shows that the classification is only weakly affected, with the decision boundaries in our fiducial model identifying a higher percentage of in-situ clusters than in the alternative model. We point out the limitations of the adopted mass loss model, particularly its inability to account for GC mass loss driven by rapidly varying tidal fields during disk crossings or interactions with molecular clouds. As highlighted by \cite{2020IAUS..351...40R}, such processes are not captured in current cosmological simulations due to insufficient spatial resolution, and their omission could significantly affect predictions of GC mass evolution.

\subsection{Orbital integrations}

To integrate the orbits of GCs, we require not only the MW gravitational potentials at different epochs (as described in section~\ref{section21}), but also the orbits of all merging satellites relative to the MW-like galaxy. This is essential because each satellite's trajectory affects the evolution of the GC system in two key ways: by perturbing the spatial distribution of the in-situ clusters, and by guiding the accreted clusters during their infall and subsequent evolution within the host galaxy.

The orbits of the satellite galaxies are also retrieved from TNG50. Hovewer, we re-integrate these merging satellite orbits into our MW potentials by taking into account dynamical friction, both to improve the time resolution of orbits that are sometimes highly jagged in the simulation due to non-linear time evolution, and to incorporate them as "moving potentials" in our full MW potential. Figure~\ref{figA0} compares the orbits of several satellites in an MW and the orbits we re-integrated for the reasons mentioned above. Between 2 and 5 Gyr, the differences between the orbits from TNG50 and our re-integrated orbits stem primarily from the fact that our approach assumes an idealized spherical distribution for both the host galaxy and its satellites, combined with an analytically modeled dynamical friction (see Figure~\ref{figA0}). In contrast, the density field in TNG50 is non-spherical, exhibiting substructures, asymmetries, and more complex local gravitational interactions. Moreover, dynamical friction in TNG50 arises intrinsically from the gravitational interactions between particles.

TNG50 provides 100 snapshots from $z=20$ and 0, distributed non-uniformly in time. In all our realizations, the satellite galaxy potentials evolve in both mass and size over time, as expected from cosmological evolution. As they orbit MW galaxies, satellites are subject to dynamical friction, which gradually alters their trajectories. The time-dependent masses and half-mass radii of the satellites are extracted directly from the TNG50 simulation and are used to compute the corresponding dynamical friction forces exerted by the MW-like host. We adopt the Chandrasekhar dynamical friction formula, as implemented in \texttt{galpy}, which closely follows the prescription of \cite{2016MNRAS.463..858P}.

For GCs, we include both dynamical friction and mass loss in the orbital integrations. At each update of the MW potential, GC mass loss is computed using the model of \cite{Kruijssen11}, which includes contributions from stellar evolution, two-body relaxation, and tidal shocks. In orbital integrations, mass loss impacts the dynamics of the GCs only in the presence of dynamical friction, which is updated at each change of potential and takes into account the mass loss of GCs. The evolution of mass and structural parameters of the host galaxy accounts for the influence of merging satellites, modeled as moving potentials, and includes dynamical friction exerted by the host galaxy on the GCs. We refer to this description as the "MW full potential + dynamical friction". For comparison, we also consider two alternative potential models: one where the MW full potential evolves over time but without including dynamical friction, and another where the MW potential is fixed without dynamical friction and corresponds to the TNG50 snapshot at $z=0$. We refer to these two configurations as "MW full potential" and "MW static potential", respectively. In our approach, we distinguish between the orbital integrations of ex-situ and in-situ GCs. Specifically, ex-situ GCs first evolve within the potential of their progenitor galaxy, which is a merging galaxy of the MW, and are later accreted by the MW, thus evolving in its potential (see Figure~\ref{fig2}). During the initial phase, ex-situ GCs orbit within the reference frame of their satellite galaxy until their orbital energy becomes positive ($E > 0$). Once this condition is met, GCs are transferred to the MW's reference frame while maintaining the moving potential of the satellite galaxy if it has not yet merged. This procedure is necessary because the calculation of dynamical friction from a galaxy is only possible within that galaxy's reference frame in \texttt{galpy}. Once accreted at a given redshift, ex-situ GCs evolve like in-situ GCs in the full MW potential. We have verified that once tagged, both ex-situ and in-situ GCs remain bound to their host galaxies ($E < 0$). We have checked that the distribution of in-situ GCs are bound to the various progenitors of the MW by computing the normalized total energy at $z = 2$.

To compute the orbits, we used a fast C integrator, dop853$_c$, implemented in \texttt{galpy}. This is an explicit Runge–Kutta method of order 8(5,3), which offers high accuracy and efficiency for integrating complex dynamical systems. Our orbital-time resolution is set to 2 Myr (500 time steps per Gyr), which is sufficient to obtain smooth and accurate orbits over these time scales. We apply this procedure to our full sample of 198 MW analogues from TNG50. Regarding the computational performance of our model, orbital integrations between $z=3$ and 0 for our entire sample take between 1 and 440 CPU hours. This variation depends on the potential description, as dynamical friction significantly slows down the code.

\section{Results}

In the following, we only consider ex-situ GCs with negative total energy at $z=0$, i.e. those that remain gravitationally bound to the MW, since we identified a small fraction of ex-situ GCs that are unbound at $z=0$, representing 2$\%$ and 10$\%$ of the ex-situ population in the static and full potential + dynamical friction models, respectively. In fact, during their escape from the satellite potential, either due to satellite evolution or merger, these GCs have sufficient energy to not be trapped within the MW potential. A smaller number of wandering GCs in the static potential is found, as expected, since it is described by the most massive MW potential at $z=0$. Then, in order to fairly compare the populations of our 198 MW analogues, which have different mass distributions as shown in the Figure~\ref{figA3}, we chose to normalize $E$-$L_{z}$ space.

In our analysis, the energy $E$ is normalized by the absolute value of the energy of a circular orbit at the stellar half-mass radius, $E_{\rm circ}(r_{\rm hm})$. $r_{\rm hm}$ provides a physically meaningful scale for two main reasons: (i) it approximately marks the transition between the in-situ and ex-situ GC populations, as in-situ clusters typically form within this radius; and (ii) this radius is directly measurable in observations, enabling robust comparisons with simulations. The angular momentum $L_z$ is normalized by $\sqrt{GM_{\rm tot}r_{\rm hm}}$, where $M_{\rm tot}$ is the total mass of the host galaxy, including DM. This normalization captures the influence of the global gravitational potential on the orbital dynamics of GCs, particularly for those on wide orbits. It also allows for meaningful comparisons between systems of different total mass while maintaining a common baryonic scale via $r_{\rm hm}$. Since the gravitational potentials used by \cite{McMillan17} and TNG50 also differ, normalizing energies and angular momenta at their stellar half-mass radius allows a fair comparison between them. In addition, varying the choice of potential model only slightly affects the classification results, provided the model remains consistent with the observational constraints, as noted by \cite{Chen24}.

We followed the dynamics of 17,726 GCs, including 13,339 in-situ and 4,387 ex-situ, across three different MW potential descriptions: MW static potential, MW full potential, and MW full potential + dynamical friction (see Section~\ref{section21} for the definitions of potentials). Thanks to our mass-loss implementation from \cite{Kruijssen11}, we can determine the final mass of the GCs, especially those that have survived. The following plots (Figure~\ref{fig3}, ~\ref{fig4} and ~\ref{fig5}) show the population of surviving GCs. In these figures, we show the limit defined by \citep{Belokurov24}. Since we aim to compare our results in the energy-angular momentum space of the MW best-fitting potential from \citet{McMillan17}, we computed the McMillan version of this limit by plotting the GCs in our normalized $E$-$L_{z}$ space and adjusting the limit so that it separates the same clusters identified as in-situ or ex-situ by \citep{Belokurov24}. On average, only 0.01$\%$ of the ex-situ GCs are destroyed, compared to 35$\%$ of the in-situ ones. This directly echoes a similar result from \cite{2017MNRAS.465.3622R}, who found that in-situ GCs experience significantly stronger tidal forces than their accreted counterparts. Assuming an initial GC mass of $10^6$ $\rm{M_{\odot}}$, we observe that, by $z=0$, the mass of surviving clusters spans a wide range from $3.2\times10^3$ $\rm{M_{\odot}}$ up to $5\times10^5$ $\rm{M_{\odot}}$. Interestingly, the ex-situ population typically retains a mass close to its initial value, highlighting their relative resilience to tidal disruption in MW-like environments.

\subsection{In-situ globular clusters}

Figure~\ref{fig3} shows that the evolution of the potential and dynamical friction have a negligible effect on the distribution of in-situ GCs in the $E$-$L_{z}$ space. As a matter of fact, the evolution of the potential will stretch some GCs toward higher energies, but most clusters are concentrated at $E < -1.2 \; |E_{\rm circ}(r_{\rm hm}^{*})|$. We observe no variation in this dense cluster region across the three descriptions. Since the potential evolves gradually, GCs can simply adjust adiabatically without significant diffusion in the $E$-$L_{z}$ plane. This is expected, as the mass growth of the MW in CDM framework occurs relatively smoothly. Regarding dynamical friction, its negligible impact is also expected, given that it is generally negligible in MW-like galaxies due to the large mass ratio between the enclosed MW mass and GC mass. Moreover, Figure~\ref{fig3} clearly shows that it is highly unlikely to find in-situ GCs at $E > -0.7\; |E_{\rm circ}(r_{\rm hm}^{*})|$.

\subsection{Ex-situ globular clusters}

Figure~\ref{fig4} depicts that the ex-situ GC distribution is more spread out in energy in the full MW potential compared to the static potential, as the evolving potential is less massive and can sustain orbits at higher energies than the MW potential at $z=0$. Dynamical friction slightly reduces the GC density distribution in the high-energy range. It also confirms the presence of an overlap between the two populations in MW-like galaxies, occurring below an energy threshold of $E < -0.7 \; |E_{\rm circ}(r_{\rm hm}^{*})|$, as ex-situ clusters span the entire energy range (see Figure~\ref{fig4}). We emphasize that ex-situ GCs are present across the full energy range, invalidating the limit proposed by \cite{Belokurov24,Massari19,2022MNRAS.513.4107C,Chen24}. In practical terms, this threshold can at best be used to exclude the in-situ population. Besides, we note that very few ex-situ clusters are found below approximately $-1.25 \; |E_{\rm circ}(r_{\rm hm}^{*})|$. 

\subsection{Comparison with Gaia MW GCs}

Figure~\ref{fig5} compares our results in the full MW potential, including dynamical friction, with the distribution of GCs in the MW as provided by Gaia \citep{Vasiliev21} in the normalized space $E$-$L_{z}$. For the observed clusters, we assumed the potential from \cite{McMillan17}, a multi-component analytic model of the MW calibrated to reproduce a wide range of observables. This MW potential has an stellar half-mass radius of 4.6 kpc. For our simulated clusters, we used a hexagonal density plot, where the color represents the dominant origin of the GCs (ex-situ or in-situ). Dominance is defined as the ratio $(N_{\rm ex-situ} - N_{\rm in-situ})/(N_{\rm ex-situ} + N_{\rm in-situ})$, where $N_{\rm ex-situ}$ and $N_{\rm in-situ}$ are the number of ex-situ and in-situ GCs, respectively. Red indicates a predominance of ex-situ clusters (+1), while blue corresponds to in-situ clusters (-1). As highlighted in previous Figures~\ref{fig3} and ~\ref{fig4}, our dominance map confirms that ex-situ clusters dominate for $E > -0.7 \;|E_{\rm circ}(r_{\rm hm}^{*})|$, while in-situ clusters dominate for $E < -1.25 \;|E_{\rm circ}(r_{\rm hm}^{*})|$ for positive $L_z$. More specifically, our density map reveals that the dominance of ex-situ clusters extends up to $-0.9 \; |E_{\rm circ}(r_{\rm hm}^{*})|$, covering a larger fraction of the observed MW GCs distribution. Our results indicate that this boundary underestimates the number of ex-situ clusters: in fact, 24$\%$ of our ex-situ clusters fall below the \cite{Belokurov24} threshold. Furthermore, if we were to adopt this boundary, our results suggest a misclassification of 10$\%$ of the clusters identified as in-situ, which are actually ex-situ. Finally, our results challenge simple energy threshold models, which we demonstrate to fail in clearly separating the two populations, and merely identifies a subset of ex-situ GCs. According to Figure~\ref{figA0}, our model tends to overestimate dynamical friction during the early orbital evolution of satellites, leading to a faster loss of orbital energy. However, the figure also shows that at the time of satellite disruption, our model overestimates the satellites’ orbital energy and consequently that of the GCs, resulting in an underestimation of the number of low-energy ex-situ objects. This implies that our boundaries should be lowered in Figure~\ref{fig5}. This further calls into question the energy threshold proposed by previous authors, which appears to exclude too large a fraction of ex-situ objects. Moreover, Figure~\ref{fig5} clearly highlights that the real challenge lies in distinguishing the two populations below this energy threshold, where an overlap of both populations persists up to -|$E_{\rm circ}(r_{\rm hm}^{*})|$. The existence of this mixing zone is further supported by \cite{Andersson2019}, who predict that $50\%$ of the GCs located in the central region of M31 were accreted.

\subsection{Our new approach}

This motivates our proposal for new boundaries that allow for the identification of a larger fraction of the ex-situ population while also capturing part of the in-situ population. We employed a Support Vector Machine (SVM)\footnote{The implementation of the SVM was done using the scikit-learn library in python.}, a supervised learning method, to classify the synthetic in-situ and ex-situ populations in the $E$-$L_{z}$ space. This algorithm aims to find an optimal hyperplane that separates the data in such a way as to maximize the margin between the two GC populations. We employed a radial basis function (RBF) kernel modeled by a squared exponential function, in order to transform the input space so that the data becomes linearly separable in a higher-dimensional feature space. The SVM was trained using this RBF kernel, with parameters chosen to minimize the classification error rate for both populations. In Figure~\ref{fig5}, the decision boundaries corresponding to probabilities of 0.05, 0.5, and 0.95 are plotted to illustrate the separation between the two populations. The 0.5 boundary results in an error of 1.35$\%$ for in-situ GCs and 11.2$\%$ for ex-situ GCs. The extreme boundary at 0.95 (0.05) excludes 99.72$\%$ (97.3$\%$) of in-situ (ex-situ) GCs while identifying 81.2$\%$ (76.4$\%$) of ex-situ (in-situ) GCs. We establish that between these two limits, distinguishing the two populations is not feasible with previous methods due to the presence of a mixing zone. The in-situ boundary approximately corresponds to an energy thresholds of $-1$. This is why, in Figure~\ref{fig5a}, where clusters are sorted from the least to the most bound in energy, Gaia MW GCs within the region of overlap are represented by squares (either in-situ (in blue) or ex-situ (in red)) without any border. We provide a new classification based on the 0.5 boundary established from the dynamics of our synthetic clusters within the full gravitational potential of 198 MW realizations. We used this threshold to identify ex-situ (in-situ) GCs, shown in red (blue) (see Figure~\ref{fig5a}). We also compare our updated classification of the MW GCs with previous studies \citep{Massari19, Malhan22, Sun23, Belokurov24, Chen24}. We would like to stress that the classifications shown in Figure 6 are taken directly from the original studies and are not recalculated by us. Since our ex-situ boundary lies at a lower energy threshold compared to \cite{Belokurov24}, Sagittarius GCs are classified as ex-situ, as expected. 

\begin{figure}[!t]
\centering
\includegraphics[width=\hsize]{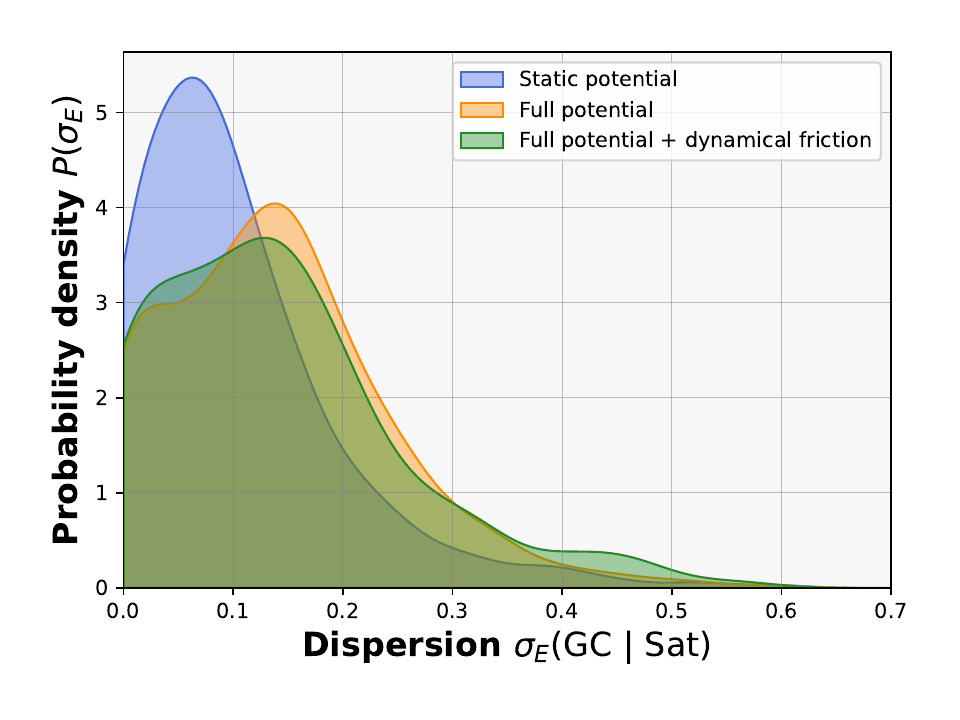}
\caption{Coherence loss in energy: Probability density P($\sigma_E$) of the energy dispersion for our ex-situ GCs in the 198 MW analogues at z=0. We calculate the energy dispersions for each ex-situ GC groups. The evolution of the galactic potential induces a greater energy dispersion for GCs originating from the same satellite.}
\label{fig8}
\end{figure}

Among the 160 MW GCs, we identify 67 as ex-situ, 38 as in-situ, and 55 within the mixing zone. This why chemical abundance could provide additional constraints to disentangle these origins, as demonstrated by \cite{2025A&A...693A.155P}. If we adopt our decision threshold (0.5 boundary), meaning that the probability of belonging to either population is equal, we find an ex-situ to in-situ number ratio close to 1.02 (79 in-situ versus 81 ex-situ). This contrasts with \cite{Belokurov24} and \cite{Chen24}, who obtain reversed ratios of approximately 0.52 (105 in-situ versus 55 ex-situ) and 0.6 (91 in-situ versus 55 ex-situ), respectively, while \cite{Sun23} reports a more balanced ratio of 1.2 (70 in-situ versus 83 ex-situ).

Figure~\ref{fig5a} clearly shows that the 33 most tightly bound GCs (with lower energy than VVV\_CL$001$) identified as in-situ are consistently found across all six studies. We also observe that our results for ex-situ GCs are in good agreement with the classifications from \cite{Massari19, Malhan22,Sun23}, except for a few clusters that were designated as in-situ in their classifications. Intriguingly, these discrepant GCs have all been identified as "Disk" GCs, meaning they exhibit properties such as low eccentricity (implying nearly circular orbits) and pericenter and apocenter radii similar to those of stars in the Galactic disk. This suggests that associating certain GCs with the disk based solely on these criteria may not be sufficiently constraining. Finally, all clusters located in the mixing zone (squares without any border in the left panel of Figure~\ref{fig5a}) have been classified as in-situ in previous studies, except for \cite{Sun23}. In their study, they identify a group of 26 GCs ('Pot') that are not associated with the bulge, the disk, or known major merger events. According to \cite{Sun23}, these clusters may originate from small accretion events and are therefore classified as ex-situ. Once again, the difficulty in distinguishing them from other GCs arises from their location in the region of overlap, where both populations overlap.

Concerning Omega Centauri (NGC 5139 in bold green), which is believed to be the nuclear star cluster of a galaxy accreted by the MW in the past, our analysis classifies it as ex-situ in contrast to most results in the literature (see Figure~\ref{fig5a}). Furthermore, \cite{Pagnini25} identified six GCs brought by the progenitor of NGC 5139 by combining Gaia EDR3 data with chemical abundances from APOGEE DR17. NGC 6752, NGC 6656, NGC 6809, NGC 6273, NGC 6205, and NGC 6254 have a fraction of stars compatible with Omega Centauri greater than 60$\%$. Our results confirm this prediction by identifying three of these clusters as ex-situ (with energies lower than or comparable to that of NGC 5139), while the remaining three lie in the mixing zone, with an uncertain but possibly ex-situ origin only for NGC 6273. Finally, we note that two of GCs identified as ex-situ (NGC 6752 and NGC 6656) are those with a compatibility greater than 80$\%$ with Omega Centauri.

\begin{figure}[!t]
\centering
\includegraphics[width=\hsize]{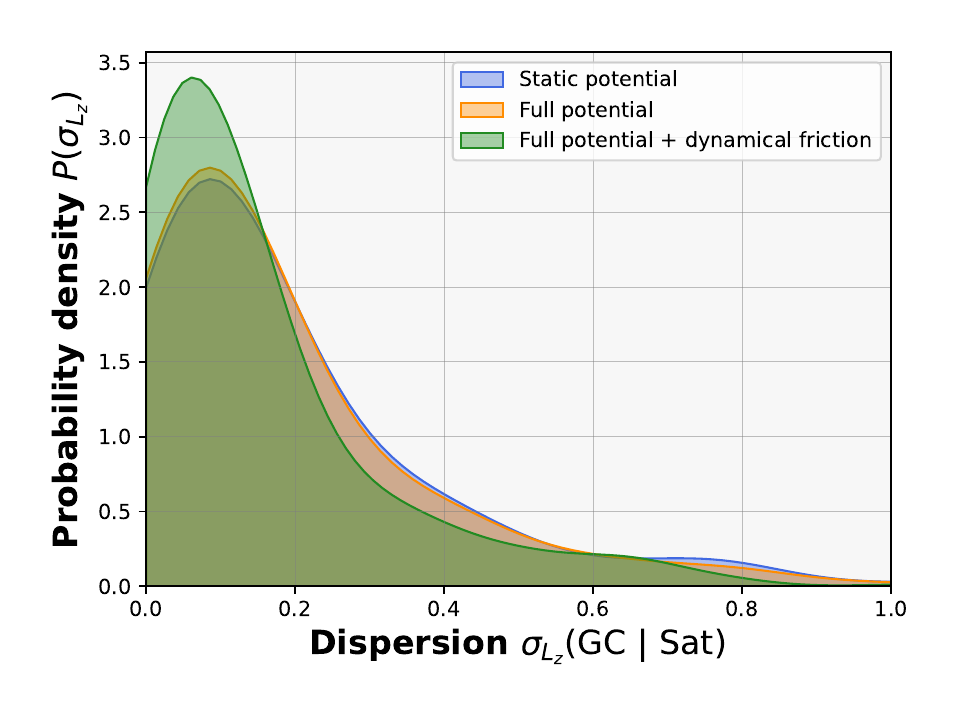}
\caption{Coherence loss in angular momentum: As in Figure~\ref{fig8}, but for the dispersion in angular momentum. Dynamical friction induces a more pronounced peak around a low $L_z$ dispersion. This indicates that a portion of the GCs tends to remain grouped in $L_z$.}
\label{fig9}
\end{figure}

Concerning another specific GC, NGC 6397, which is the second closest GC to the Sun, it is a low-mass and very compact cluster. Due to its proximity to the MW center, it should have experienced numerous tidal interactions, making it a strong candidate for the development of tidal tails. Moreover, \cite{Balbinot18} estimated that it should have lost 72$\%$ of its initial mass. Actually, simulations of NGC 6397 generally predict prominent and extended tidal tails (kpc scale) around this cluster \citep{Boldrini21,Arnold25}. However, the existence of tidal tails around NGC 6397 remains debated. \cite{2000A&A...359..907L} reported overdensities that were classified as unreliable due to uncertainties in the dust distribution around the cluster. More recently, \cite{Ibata24} detected a possible tail extending over 18 degrees on the sky using Gaia data, but these findings were challenged by \cite{Boldrini21}, who did not detect any tidal tail. Ultimately, identifying potential stream members of NGC 6397 in Gaia data with precision remains a complex challenge. This is crucial to either confirm the absence of a stream, the presence of a sub-kpc stream, or the existence of a prominent stream as predicted by simulations assuming an in-situ evolution. Our results classify NGC 6397 (which has an energy level comparable to that of Omega Centauri) as in-situ, but this classification remains highly uncertain given its position near the 0.5 boundary (see Figure~\ref{fig5}). Therefore, this does not rule out scenarios where either no tidal stream is present or only a sub-kiloparsec stream exists. Indeed, this aligns with the hypothesis proposed by \cite{Boldrini21} that one possible ex-situ scenario for this cluster is that it was initially embedded in a DM minihalo. In this scenario, the minihalo could have acted as a protective envelope, preventing the tidal stripping of stars while being gradually disrupted and removed over the course of the cluster’s evolution. The combination of future releases from Gaia and Euclid will provide a better characterization of the tidal features of this cluster and further constrain its origin \citep{Massari24}.

Finally, there is a group of 23 GCs whose origin remains unidentified but which have been classified as 'low-energy' GCs by \cite{Massari19} due to their low energy and near-zero angular momentum. Our results indicate that this population has energies lying precisely in the mixing zone (see GCs in grey in Figure~\ref{fig5a}). The fact that these clusters fall within this energy range further complicates the determination of their origin and, more importantly, suggests the absence of a common progenitor. Besides, we find that 10 of them have an ex-situ origin.

\subsection{$E$-$L_{z}$ decoherence of ex-situ GC groups}

Our final investigation focuses on the distribution of ex-situ GC groups originating from the same satellite, still in phase space $E$-$L_{z}$, as shown in Figure~\ref{fig7}, for our three descriptions of the MW potential over time (see Section 2.1 for definitions of potentials). An important result is that GCs from the same satellite can have different energies, as well as both positive and negative z-components of angular momentum. Furthermore, the boundary proposed by \cite{Belokurov24} can separate GCs originating from the same satellite. Upon visual inspection of Figure~\ref{fig7}, we observe that the evolving potential tends to break the coherence of GC groups from the same satellite across five MW galaxies in our sample, whose orbits are provided in Figure~\ref{fig6}. This decoherence manifests as a diffusion of GC groups towards higher energies, due to the changes in the potential experienced by the ex-situ GCs. Unlike in-situ GCs, ex-situ GCs take longer to adapt to variations in the gravitational potential, resulting in an increase in their energy dispersion. Interestingly, the addition of dynamical friction tends to accelerate this adaptation process, as we observe a slight decrease in the energies of the GCs. A notable effect initiated by dynamical friction is its tendency to align the angular momentum of GCs from the same satellite. Indeed, dynamical friction is a process that removes angular momentum and orbital energy, driving the GCs towards more radial orbits.

To confirm these behaviors across our sample of 198 MWs, we calculate the energy and angular momentum dispersions for each ex-situ GC group at $z=0$, as shown in Figures~\ref{fig8} and~\ref{fig9}, respectively. In Figure~\ref{fig8}, for the static potential case (blue), we observe a pronounced peak towards small dispersions. When we account for an evolving potential (orange), the distribution shifts slightly towards higher values, with a flattening of the main peak. This suggests that the evolution of the galactic potential induces a greater energy dispersion for GCs originating from the same satellite. This confirms that variations in the potential well alter the orbits of accreted GCs, as visually observed in the example in Figure~\ref{fig6}. The addition of dynamical friction (green) slightly reduces the energy dispersion. As seen in Figure~\ref{fig8}, the progressive energy loss of the GCs due to dynamical friction slightly narrows the energy gap between them. In Figure~\ref{fig9}, the inclusion of an evolving potential does not affect the $L_z$ distribution of ex-situ GCs at $z=0$, whereas dynamical friction induces a more pronounced peak around a low $L_z$ dispersion. This indicates that a portion of the GCs tends to remain grouped in $L_z$. Once again, this is expected, as dynamical friction removes angular momentum from the orbiting GCs.

\section{Caveats}

In this work, as a first approximation, we tagged ex-situ GCs at $z=3$ in satellite galaxies and in-situ GCs at $z=2$ in the MW progenitors. To increase realism, future studies will need to tag clusters at all redshifts, based, for example, on a formation efficiency parameter derived from gas and stellar quantities resolved in the simulation (e.g., as in \cite{Pfeffer18, ReinaCampos22, Grudic23, Newton24, Rodriguez23}). However, high-resolution or zoom-in simulations, which typically generate a maximum of 20 MW-like galaxies, are required to calculate these parameters because a gas resolution of the order of 1-10 pc is necessary to obtain reliable information on gas pressure and star formation, which enhance the formation of GCs. Although TNG50 is not as precise, with an average cell size of 70–140 pc in the star-forming regions of galaxies, it allows us to access approximately 200 MW-like galaxies \citep{Pillepich19}. Although our approach is statistically consistent with the estimated ages of all GCs in the MW, it is not consistent with younger clusters with lower masses ($10^{4-5}$ M${\sun}$). Moreover, tagging clusters at lower redshifts will increase the population of clusters that survive in environments like those of MW-type galaxies. In the same spirit, a tagging method based on a formation efficiency parameter using gas and stellar quantities would provide a more realistic mass for the GCs. However, the goal of our initial approach was to model the average behavior of clusters with masses on the order of $10^6$ M${\sun}$. In our analysis, we only consider satellite galaxies with stellar masses above $10^7$ $\rm{M_{\odot}}$. However, lower-mass galaxies, which are not resolved in this study, may be highly efficient sites of GC formation, as indicated by recent results from the EDGE simulations \citep{Taylor25}. This limitation clearly motivates future investigations using zoom-in simulations that can better resolve low-mass DM halos and their contribution to the GC population.

Another point to improve our approach is to include the presence of the MW's disk, particularly to account for its dynamical impact on the in-situ population. In this work, our MWs are only modeled with DM and stellar spherical components. Although these components have been studied in TNG50, they were not cataloged in the same way as the other components for our MW sample to be incorporated into our study \citep{Sotillo22}. To quantify the impact of the disk on our GC classification, we added a fixed Miyamoto–Nagai potential representing the observed disk at $z=0$ to all our TNG50 MWs. This test does not aim to provide a realistic modelling of the disk's evolution, but rather serves as an upper-limit case to assess how much our results could be affected by this underestimation of tidal disruption caused by the absence of disk in our modelling. The table~\ref{tab1} shows that both the SVM boundaries and the classification remain largely unaffected by this extreme scenario. This can be explained by two main points: (1) the disk primarily affects the in-situ population, and (2) our classification relies on GC survivability rather than their final mass. In fact, we found that in-situ GCs lose more mass in the presence of the fixed disk. An alternative to avoid imposing fixed structures (Hernquist bulge, exponential disk, NFW DM halo) and breaking this idealized modeling of the MW via a spherical potential is to use methods like \texttt{AGAMA} \citep{AGAMA}. This would allow us to build a self-consistent galactic potential based on the galaxy's full distribution function, including non-trivial effects such as local density variations.

\section{Conclusion}

In this work, we presented a new cosmological post-processing framework to study the hierarchical assembly of GCs in MW-like galaxies. By coupling the cosmological hydrodynamical simulation TNG50 with orbital integrations using \texttt{galpy}, and by including realistic treatments of dynamical friction and mass loss, we followed the evolution of 18,000 GCs — both in-situ and ex-situ — across 198 MW analogues from redshift $z=3$ to $z=0$. This approach allowed us to explore, for the first time at this statistical scale, how these two GC populations dynamically evolve and mix within time-evolving galactic potentials shaped by mergers and mass accretion.

We find that in-situ GCs predominantly occupy a well-defined region of the normalized energy–angular momentum $E$-$L_{z}$ space, remaining tightly bound and dynamically cold even in the presence of an evolving MW potential. In contrast, ex-situ GCs span a broader range in energy and angular momentum, reflecting their varied accretion histories. Importantly, we confirm the existence of a significant mixing zone below the energy threshold $E < -0.7 \; |E_{\rm circ}(r_{\rm hm}^{*})|$, where both in-situ and ex-situ GCs overlap. This undermines previous attempts to cleanly separate the two populations using only the integral of motions via a single cut in energy, which based on our models will not be able to reproduce a large fraction of ex-situ clusters and will misclassify many in-situ ones.

We proposed a new classification scheme, which allows a probabilistic separation of the two populations and highlights the limitations of binary thresholds. Applying this scheme to the observed MW GC system using Gaia kinematics, we identify 81 GCs as ex-situ, 79 as in-situ (see Figures~\ref{fig5} and ~\ref{fig5a}). This revised classification leads to a more balanced in-situ to ex-situ ratio. Notably, we confirm the ex-situ nature of Omega Centauri and its associated group of GCs, as well as the Sagittarius GCs, and identify potential misclassifications among clusters previously associated with the Galactic disk.

Explicit modeling of mass loss during orbital integration, using the prescription of \cite{Kruijssen11}, allowed us to find that in-situ GCs experience significantly stronger tidal disruption, leading to an average survival fraction of only $\sim 65\%$, while ex-situ GCs are more resilient, with a destruction rate below $1\%$. As a result, surviving ex-situ clusters retain masses closer to their initial values, while many in-situ clusters are heavily stripped, in agreement with \cite{2017MNRAS.465.3622R}. This asymmetry in mass loss between in-situ and ex-situ GCs may provide an additional observational handle to disentangle the two populations. It could also offer a possible explanation for the extremely metal-poor stellar streams recently identified by \citet{Ibata24}, which have not been linked to any known GC progenitor. Our results suggest a plausible scenario in which such streams may represent the remnants of a disrupted, ancient population of in-situ GCs, destroyed during the early phases of the Galaxy's evolution. We emphasize that this is not a firm prediction, but rather a possible interpretation consistent with the asymmetry in GC mass loss observed in our model.

Furthermore, we studied the loss of orbital coherence among GC groups accreted from the same satellite. We find that the evolving MW potential, combined with dynamical friction, progressively erases the initial correlations in $E$-$L_{z}$ space. This challenges the assumption that GCs accreted from the same progenitor should remain kinematically coherent, even in action space, and calls for caution when associating groups of clusters with past accretion events.

Overall, our results show that the dynamical separation of GC populations requires modeling their full cosmological context, including satellite orbits, potential evolution, and external GC evolution. Our method bridges the gap between large-scale statistical studies and the dynamical modeling of individual clusters, offering new insights into the assembly history of the MW. Future improvements will involve tagging GCs based on more physically motivated formation efficiencies, including baryonic structures like the MW disk, and using higher-resolution MW simulations such as VINTERGATAN \citep{VINTERGATAN} and HESTIA \cite{Hestia} to refine GC initial conditions. Such developments will further enhance our ability to disentangle the complex origin of the Galactic GC system in the Gaia era, as well as with Euclid \citep{Voggel25,Sai2025V1,Sai2025V2}.

\section{Data Availability}

The data underlying this article is available through reasonable request to the author. The code and GC data will be available at the following URL: \href{https://github.com/Blackholan}{https://github.com/Blackholan.}

\section{Acknowledgements}

PB acknowledges funding from the CNES post-doctoral fellowship program. PB, PDM and GP are grateful to the "Action Thématique de Cosmologie et Galaxies (ATCG), Programme National ASTRO of the INSU (Institut National des Sciences de l'Univers) for supporting this research, in the framework of the project "Coevolution of globular clusters and dwarf galaxies, in the context of hierarchical galaxy formation: from the Milky Way to the nearby Universe" (PI: A. Lançon). PB thanks members of the Euclid working group, "Extragalactic GCs", for contributing with useful comments in the early stages of this project. PB wants to express his gratitude to Misha Haywood for suggesting a more impactful title. The IllustrisTNG simulations were undertaken with compute time awarded by the Gauss Centre for Supercomputing (GCS) under GCS Large-Scale Projects GCS-ILLU and GCS-DWAR on the GCS share of the supercomputer Hazel Hen at the High Performance Computing Center Stuttgart (HLRS), as well as on the machines of the Max Planck Computing and Data Facility (MPCDF) in Garching, Germany.


\bibliography{src}

\begin{thebibliography}{103}
\expandafter\ifx\csname natexlab\endcsname\relax\def\natexlab#1{#1}\fi

\bibitem[{{Agertz} {et~al.}(2021){Agertz}, {Renaud}, {Feltzing}, {Read}, {Ryde}, {Andersson}, {Rey}, {Bensby}, \& {Feuillet}}]{VINTERGATAN}
{Agertz}, O., {Renaud}, F., {Feltzing}, S., {et~al.} 2021, \mnras, 503, 5826

\bibitem[{{Amarante} {et~al.}(2022){Amarante}, {Debattista}, {Beraldo e Silva}, {Laporte}, \& {Deg}}]{2022ApJ...937...12A}
{Amarante}, J. A.~S., {Debattista}, V.~P., {Beraldo e Silva}, L., {Laporte}, C. F.~P., \& {Deg}, N. 2022, \apj, 937, 12

\bibitem[{{Andersson} \& {Davies}(2019)}]{Andersson2019}
{Andersson}, E.~P. \& {Davies}, M.~B. 2019, \mnras, 485, 4134

\bibitem[{{Andersson} {et~al.}(2024){Andersson}, {Mac Low}, {Agertz}, {Renaud}, \& {Li}}]{Andersson24}
{Andersson}, E.~P., {Mac Low}, M.-M., {Agertz}, O., {Renaud}, F., \& {Li}, H. 2024, \aap, 681, A28

\bibitem[{{Arnold} \& {Baumgardt}(2025)}]{Arnold25}
{Arnold}, A.~D. \& {Baumgardt}, H. 2025, \mnras, 537, 1807

\bibitem[{{Balbinot} \& {Gieles}(2018)}]{Balbinot18}
{Balbinot}, E. \& {Gieles}, M. 2018, \mnras, 474, 2479

\bibitem[{{Baumgardt} {et~al.}(2019){Baumgardt}, {Hilker}, {Sollima}, \& {Bellini}}]{Baumgardt19}
{Baumgardt}, H., {Hilker}, M., {Sollima}, A., \& {Bellini}, A. 2019, \mnras, 482, 5138

\bibitem[{{Beasley}(2020)}]{Beasley20}
{Beasley}, M.~A. 2020, in Reviews in Frontiers of Modern Astrophysics; From Space Debris to Cosmology, ed. P.~{Kab{\'a}th}, D.~{Jones}, \& M.~{Skarka}, 245--277

\bibitem[{{Bekki} \& {Freeman}(2003)}]{2003MNRAS.346L..11B}
{Bekki}, K. \& {Freeman}, K.~C. 2003, \mnras, 346, L11

\bibitem[{{Belokurov} \& {Kravtsov}(2023)}]{Belokurov23}
{Belokurov}, V. \& {Kravtsov}, A. 2023, \mnras, 525, 4456

\bibitem[{{Belokurov} \& {Kravtsov}(2024)}]{Belokurov24}
{Belokurov}, V. \& {Kravtsov}, A. 2024, \mnras, 528, 3198

\bibitem[{{Binney} \& {Tremaine}(2008)}]{2008gady.book.....B}
{Binney}, J. \& {Tremaine}, S. 2008, {Galactic Dynamics: Second Edition}

\bibitem[{{Boldrini} \& {Vitral}(2021)}]{Boldrini21}
{Boldrini}, P. \& {Vitral}, E. 2021, \mnras, 507, 1814

\bibitem[{{Bovy}(2015)}]{Bovy15}
{Bovy}, J. 2015, \apjs, 216, 29

\bibitem[{{Bullock} \& {Johnston}(2005)}]{Bullock05}
{Bullock}, J.~S. \& {Johnston}, K.~V. 2005, \apj, 635, 931

\bibitem[{{Burkert} \& {Forbes}(2020)}]{Burkert20}
{Burkert}, A. \& {Forbes}, D.~A. 2020, \aj, 159, 56

\bibitem[{{Callingham} {et~al.}(2022){Callingham}, {Cautun}, {Deason}, {Frenk}, {Grand}, \& {Marinacci}}]{2022MNRAS.513.4107C}
{Callingham}, T.~M., {Cautun}, M., {Deason}, A.~J., {et~al.} 2022, \mnras, 513, 4107

\bibitem[{{Carraro} \& {Lia}(2000)}]{2000A&A...357..977C}
{Carraro}, G. \& {Lia}, C. 2000, \aap, 357, 977

\bibitem[{{Chen} \& {Gnedin}(2023)}]{Chen23}
{Chen}, Y. \& {Gnedin}, O.~Y. 2023, \mnras, 522, 5638

\bibitem[{{Chen} \& {Gnedin}(2024)}]{Chen24}
{Chen}, Y. \& {Gnedin}, O.~Y. 2024, The Open Journal of Astrophysics, 7, 23

\bibitem[{{Choksi} \& {Gnedin}(2019)}]{2019MNRAS.488.5409C}
{Choksi}, N. \& {Gnedin}, O.~Y. 2019, \mnras, 488, 5409

\bibitem[{{Cooper} {et~al.}(2010){Cooper}, {Cole}, {Frenk}, {White}, {Helly}, {Benson}, {De Lucia}, {Helmi}, {Jenkins}, {Navarro}, {Springel}, \& {Wang}}]{Cooper10}
{Cooper}, A.~P., {Cole}, S., {Frenk}, C.~S., {et~al.} 2010, \mnras, 406, 744

\bibitem[{{Creasey} {et~al.}(2019){Creasey}, {Sales}, {Peng}, \& {Sameie}}]{Creasey19}
{Creasey}, P., {Sales}, L.~V., {Peng}, E.~W., \& {Sameie}, O. 2019, \mnras, 482, 219

\bibitem[{{Deng} {et~al.}(2024){Deng}, {Li}, {Liu}, {Kannan}, {Smith}, \& {Bryan}}]{Deng24}
{Deng}, Y., {Li}, H., {Liu}, B., {et~al.} 2024, \aap, 691, A231

\bibitem[{{Di Matteo} {et~al.}(2019){Di Matteo}, {Haywood}, {Lehnert}, {Katz}, {Khoperskov}, {Snaith}, {G{\'o}mez}, \& {Robichon}}]{2019A&A...632A...4D}
{Di Matteo}, P., {Haywood}, M., {Lehnert}, M.~D., {et~al.} 2019, \aap, 632, A4

\bibitem[{{Diemand} {et~al.}(2007){Diemand}, {Kuhlen}, \& {Madau}}]{2007ApJ...657..262D}
{Diemand}, J., {Kuhlen}, M., \& {Madau}, P. 2007, \apj, 657, 262

\bibitem[{{Diemer} {et~al.}(2013){Diemer}, {More}, \& {Kravtsov}}]{Diemer13}
{Diemer}, B., {More}, S., \& {Kravtsov}, A.~V. 2013, \apj, 766, 25

\bibitem[{{Doppel} {et~al.}(2023){Doppel}, {Sales}, {Nelson}, {Pillepich}, {Abadi}, {Peng}, {Marinacci}, {Naiman}, {Torrey}, {Vogelsberger}, {Weinberger}, \& {Hernquist}}]{Doppel23}
{Doppel}, J.~E., {Sales}, L.~V., {Nelson}, D., {et~al.} 2023, \mnras, 518, 2453

\bibitem[{{Dubois} {et~al.}(2021){Dubois}, {Beckmann}, {Bournaud}, {Choi}, {Devriendt}, {Jackson}, {Kaviraj}, {Kimm}, {Kraljic}, {Laigle}, {Martin}, {Park}, {Peirani}, {Pichon}, {Volonteri}, \& {Yi}}]{Dubois21}
{Dubois}, Y., {Beckmann}, R., {Bournaud}, F., {et~al.} 2021, \aap, 651, A109

\bibitem[{{Forbes} \& {Bridges}(2010)}]{Forbes10}
{Forbes}, D.~A. \& {Bridges}, T. 2010, \mnras, 404, 1203

\bibitem[{{Gaia Collaboration} {et~al.}(2021){Gaia Collaboration}, {Brown}, {Vallenari}, {Prusti}, {de Bruijne}, {Babusiaux}, {Biermann}, {Creevey}, {Evans}, {Eyer}, {Hutton}, {Jansen}, {Jordi}, {Klioner}, {Lammers}, {Lindegren}, {Luri}, {Mignard}, {Panem}, {Pourbaix}, {Randich}, {Sartoretti}, {Soubiran}, {Walton}, {Arenou}, {Bailer-Jones}, {Bastian}, {Cropper}, {Drimmel}, {Katz}, {Lattanzi}, {van Leeuwen}, {Bakker}, {Cacciari}, {Casta{\~n}eda}, {De Angeli}, {Ducourant}, {Fabricius}, {Fouesneau}, {Fr{\'e}mat}, {Guerra}, {Guerrier}, {Guiraud}, {Jean-Antoine Piccolo}, {Masana}, {Messineo}, {Mowlavi}, {Nicolas}, {Nienartowicz}, {Pailler}, {Panuzzo}, {Riclet}, {Roux}, {Seabroke}, {Sordo}, {Tanga}, {Th{\'e}venin}, {Gracia-Abril}, {Portell}, {Teyssier}, {Altmann}, {Andrae}, {Bellas-Velidis}, {Benson}, {Berthier}, {Blomme}, {Brugaletta}, {Burgess}, {Busso}, {Carry}, {Cellino}, {Cheek}, {Clementini}, {Damerdji}, {Davidson}, {Delchambre}, {Dell'Oro}, {Fern{\'a}ndez-Hern{\'a}ndez}, {Galluccio}, {Garc{\'\i}a-Lario},
  {Garcia-Reinaldos}, {Gonz{\'a}lez-N{\'u}{\~n}ez}, {Gosset}, {Haigron}, {Halbwachs}, {Hambly}, {Harrison}, {Hatzidimitriou}, {Heiter}, {Hern{\'a}ndez}, {Hestroffer}, {Hodgkin}, {Holl}, {Jan{\ss}en}, {Jevardat de Fombelle}, {Jordan}, {Krone-Martins}, {Lanzafame}, {L{\"o}ffler}, {Lorca}, {Manteiga}, {Marchal}, {Marrese}, {Moitinho}, {Mora}, {Muinonen}, {Osborne}, {Pancino}, {Pauwels}, {Petit}, {Recio-Blanco}, {Richards}, {Riello}, {Rimoldini}, {Robin}, {Roegiers}, {Rybizki}, {Sarro}, {Siopis}, {Smith}, {Sozzetti}, {Ulla}, {Utrilla}, {van Leeuwen}, {van Reeven}, {Abbas}, {Abreu Aramburu}, {Accart}, {Aerts}, {Aguado}, {Ajaj}, {Altavilla}, {{\'A}lvarez}, {{\'A}lvarez Cid-Fuentes}, {Alves}, {Anderson}, {Anglada Varela}, {Antoja}, {Audard}, {Baines}, {Baker}, {Balaguer-N{\'u}{\~n}ez}, {Balbinot}, {Balog}, {Barache}, {Barbato}, {Barros}, {Barstow}, {Bartolom{\'e}}, {Bassilana}, {Bauchet}, {Baudesson-Stella}, {Becciani}, {Bellazzini}, {Bernet}, {Bertone}, {Bianchi}, {Blanco-Cuaresma}, {Boch}, {Bombrun}, {Bossini},
  {Bouquillon}, {Bragaglia}, {Bramante}, {Breedt}, {Bressan}, {Brouillet}, {Bucciarelli}, {Burlacu}, {Busonero}, {Butkevich}, {Buzzi}, {Caffau}, {Cancelliere}, {C{\'a}novas}, {Cantat-Gaudin}, {Carballo}, {Carlucci}, {Carnerero}, {Carrasco}, {Casamiquela}, {Castellani}, {Castro-Ginard}, {Castro Sampol}, {Chaoul}, {Charlot}, {Chemin}, {Chiavassa}, {Cioni}, {Comoretto}, {Cooper}, {Cornez}, {Cowell}, {Crifo}, {Crosta}, {Crowley}, {Dafonte}, {Dapergolas}, {David}, \& {David}}]{Gaia21}
{Gaia Collaboration}, {Brown}, A.~G.~A., {Vallenari}, A., {et~al.} 2021, \aap, 649, A1

\bibitem[{{Grudi{\'c}} {et~al.}(2023){Grudi{\'c}}, {Hafen}, {Rodriguez}, {Guszejnov}, {Lamberts}, {Wetzel}, {Boylan-Kolchin}, \& {Faucher-Gigu{\`e}re}}]{Grudic23}
{Grudi{\'c}}, M.~Y., {Hafen}, Z., {Rodriguez}, C.~L., {et~al.} 2023, \mnras, 519, 1366

\bibitem[{{Halbesma} {et~al.}(2020){Halbesma}, {Grand}, {G{\'o}mez}, {Marinacci}, {Pakmor}, {Trick}, {Busch}, \& {White}}]{Halbesma20}
{Halbesma}, T. L.~R., {Grand}, R. J.~J., {G{\'o}mez}, F.~A., {et~al.} 2020, \mnras, 496, 638

\bibitem[{{Helmi}(2020)}]{Helmi20}
{Helmi}, A. 2020, \araa, 58, 205

\bibitem[{{Hernquist}(1990)}]{Hernquist90}
{Hernquist}, L. 1990, \apj, 356, 359

\bibitem[{{Ibata} {et~al.}(2024){Ibata}, {Malhan}, {Tenachi}, {Ardern-Arentsen}, {Bellazzini}, {Bianchini}, {Bonifacio}, {Caffau}, {Diakogiannis}, {Errani}, {Famaey}, {Ferrone}, {Martin}, {di Matteo}, {Monari}, {Renaud}, {Starkenburg}, {Thomas}, {Viswanathan}, \& {Yuan}}]{Ibata24}
{Ibata}, R., {Malhan}, K., {Tenachi}, W., {et~al.} 2024, \apj, 967, 89

\bibitem[{{Jean-Baptiste} {et~al.}(2017){Jean-Baptiste}, {Di Matteo}, {Haywood}, {G{\'o}mez}, {Montuori}, {Combes}, \& {Semelin}}]{2017A&A...604A.106J}
{Jean-Baptiste}, I., {Di Matteo}, P., {Haywood}, M., {et~al.} 2017, \aap, 604, A106

\bibitem[{{Khoperskov} {et~al.}(2023{\natexlab{a}}){Khoperskov}, {Minchev}, {Libeskind}, {Haywood}, {Di Matteo}, {Belokurov}, {Steinmetz}, {Gomez}, {Grand}, {Hoffman}, {Knebe}, {Sorce}, {Spaare}, {Tempel}, \& {Vogelsberger}}]{2023A&A...677A..89K}
{Khoperskov}, S., {Minchev}, I., {Libeskind}, N., {et~al.} 2023{\natexlab{a}}, \aap, 677, A89

\bibitem[{{Khoperskov} {et~al.}(2023{\natexlab{b}}){Khoperskov}, {Minchev}, {Libeskind}, {Haywood}, {Di Matteo}, {Belokurov}, {Steinmetz}, {Gomez}, {Grand}, {Hoffman}, {Knebe}, {Sorce}, {Spaare}, {Tempel}, \& {Vogelsberger}}]{2023A&A...677A..90K}
{Khoperskov}, S., {Minchev}, I., {Libeskind}, N., {et~al.} 2023{\natexlab{b}}, \aap, 677, A90

\bibitem[{{Kim} {et~al.}(2018){Kim}, {Ma}, {Grudi{\'c}}, {Hopkins}, {Hayward}, {Wetzel}, {Faucher-Gigu{\`e}re}, {Kere{\v{s}}}, {Garrison-Kimmel}, \& {Murray}}]{Kim18}
{Kim}, J.-h., {Ma}, X., {Grudi{\'c}}, M.~Y., {et~al.} 2018, \mnras, 474, 4232

\bibitem[{{Kravtsov} \& {Gnedin}(2005)}]{Kravtsov05}
{Kravtsov}, A.~V. \& {Gnedin}, O.~Y. 2005, \apj, 623, 650

\bibitem[{{Kruijssen} {et~al.}(2011){Kruijssen}, {Pelupessy}, {Lamers}, {Portegies Zwart}, \& {Icke}}]{Kruijssen11}
{Kruijssen}, J.~M.~D., {Pelupessy}, F.~I., {Lamers}, H. J.~G.~L.~M., {Portegies Zwart}, S.~F., \& {Icke}, V. 2011, \mnras, 414, 1339

\bibitem[{{Kruijssen} {et~al.}(2020){Kruijssen}, {Pfeffer}, {Chevance}, {Bonaca}, {Trujillo-Gomez}, {Bastian}, {Reina-Campos}, {Crain}, \& {Hughes}}]{Kruijssen20}
{Kruijssen}, J.~M.~D., {Pfeffer}, J.~L., {Chevance}, M., {et~al.} 2020, \mnras, 498, 2472

\bibitem[{{Kruijssen} {et~al.}(2019{\natexlab{a}}){Kruijssen}, {Pfeffer}, {Crain}, \& {Bastian}}]{Kruijssen19E}
{Kruijssen}, J.~M.~D., {Pfeffer}, J.~L., {Crain}, R.~A., \& {Bastian}, N. 2019{\natexlab{a}}, \mnras, 486, 3134

\bibitem[{{Kruijssen} {et~al.}(2019{\natexlab{b}}){Kruijssen}, {Pfeffer}, {Reina-Campos}, {Crain}, \& {Bastian}}]{Kruijssen19}
{Kruijssen}, J.~M.~D., {Pfeffer}, J.~L., {Reina-Campos}, M., {Crain}, R.~A., \& {Bastian}, N. 2019{\natexlab{b}}, \mnras, 486, 3180

\bibitem[{{Lah{\'e}n} {et~al.}(2020){Lah{\'e}n}, {Naab}, {Johansson}, {Elmegreen}, {Hu}, {Walch}, {Steinwandel}, \& {Moster}}]{Lahen20}
{Lah{\'e}n}, N., {Naab}, T., {Johansson}, P.~H., {et~al.} 2020, \apj, 891, 2

\bibitem[{{Laporte} {et~al.}(2013){Laporte}, {White}, {Naab}, \& {Gao}}]{Laporte13}
{Laporte}, C. F.~P., {White}, S. D.~M., {Naab}, T., \& {Gao}, L. 2013, \mnras, 435, 901

\bibitem[{{Leaman} {et~al.}(2013){Leaman}, {VandenBerg}, \& {Mendel}}]{Leaman2013}
{Leaman}, R., {VandenBerg}, D.~A., \& {Mendel}, J.~T. 2013, \mnras, 436, 122

\bibitem[{{Lee} {et~al.}(1999){Lee}, {Joo}, {Sohn}, {Rey}, {Lee}, \& {Walker}}]{1999Natur.402...55L}
{Lee}, Y.~W., {Joo}, J.~M., {Sohn}, Y.~J., {et~al.} 1999, \nat, 402, 55

\bibitem[{{Leon} {et~al.}(2000){Leon}, {Meylan}, \& {Combes}}]{2000A&A...359..907L}
{Leon}, S., {Meylan}, G., \& {Combes}, F. 2000, \aap, 359, 907

\bibitem[{{Li} {et~al.}(2017){Li}, {Gnedin}, {Gnedin}, {Meng}, {Semenov}, \& {Kravtsov}}]{Li17}
{Li}, H., {Gnedin}, O.~Y., {Gnedin}, N.~Y., {et~al.} 2017, \apj, 834, 69

\bibitem[{{Li} {et~al.}(2022){Li}, {Vogelsberger}, {Bryan}, {Marinacci}, {Sales}, \& {Torrey}}]{Li22}
{Li}, H., {Vogelsberger}, M., {Bryan}, G.~L., {et~al.} 2022, \mnras, 514, 265

\bibitem[{{Libeskind} {et~al.}(2020){Libeskind}, {Carlesi}, {Grand}, {Khalatyan}, {Knebe}, {Pakmor}, {Pilipenko}, {Pawlowski}, {Sparre}, {Tempel}, {Wang}, {Courtois}, {Gottl{\"o}ber}, {Hoffman}, {Minchev}, {Pfrommer}, {Sorce}, {Springel}, {Steinmetz}, {Tully}, {Vogelsberger}, \& {Yepes}}]{Hestia}
{Libeskind}, N.~I., {Carlesi}, E., {Grand}, R. J.~J., {et~al.} 2020, \mnras, 498, 2968

\bibitem[{{Lux} {et~al.}(2010){Lux}, {Read}, \& {Lake}}]{2010MNRAS.406.2312L}
{Lux}, H., {Read}, J.~I., \& {Lake}, G. 2010, \mnras, 406, 2312

\bibitem[{{Ma} {et~al.}(2020){Ma}, {Grudi{\'c}}, {Quataert}, {Hopkins}, {Faucher-Gigu{\`e}re}, {Boylan-Kolchin}, {Wetzel}, {Kim}, {Murray}, \& {Kere{\v{s}}}}]{Ma20}
{Ma}, X., {Grudi{\'c}}, M.~Y., {Quataert}, E., {et~al.} 2020, \mnras, 493, 4315

\bibitem[{{Mackereth} \& {Bovy}(2020)}]{2020MNRAS.492.3631M}
{Mackereth}, J.~T. \& {Bovy}, J. 2020, \mnras, 492, 3631

\bibitem[{{Mackey} \& {van den Bergh}(2005)}]{Mackey2005}
{Mackey}, A.~D. \& {van den Bergh}, S. 2005, \mnras, 360, 631

\bibitem[{{Majewski} {et~al.}(2000){Majewski}, {Patterson}, {Dinescu}, {Johnson}, {Ostheimer}, {Kunkel}, \& {Palma}}]{2000LIACo..35..619M}
{Majewski}, S.~R., {Patterson}, R.~J., {Dinescu}, D.~I., {et~al.} 2000, in Liege International Astrophysical Colloquia, Vol.~35, Liege International Astrophysical Colloquia, ed. A.~{Noels}, P.~{Magain}, D.~{Caro}, E.~{Jehin}, G.~{Parmentier}, \& A.~A. {Thoul}, 619

\bibitem[{{Majewski} {et~al.}(2017){Majewski}, {Schiavon}, {Frinchaboy}, {Allende Prieto}, {Barkhouser}, {Bizyaev}, {Blank}, {Brunner}, {Burton}, {Carrera}, {Chojnowski}, {Cunha}, {Epstein}, {Fitzgerald}, {Garc{\'\i}a P{\'e}rez}, {Hearty}, {Henderson}, {Holtzman}, {Johnson}, {Lam}, {Lawler}, {Maseman}, {M{\'e}sz{\'a}ros}, {Nelson}, {Nguyen}, {Nidever}, {Pinsonneault}, {Shetrone}, {Smee}, {Smith}, {Stolberg}, {Skrutskie}, {Walker}, {Wilson}, {Zasowski}, {Anders}, {Basu}, {Beland}, {Blanton}, {Bovy}, {Brownstein}, {Carlberg}, {Chaplin}, {Chiappini}, {Eisenstein}, {Elsworth}, {Feuillet}, {Fleming}, {Galbraith-Frew}, {Garc{\'\i}a}, {Garc{\'\i}a-Hern{\'a}ndez}, {Gillespie}, {Girardi}, {Gunn}, {Hasselquist}, {Hayden}, {Hekker}, {Ivans}, {Kinemuchi}, {Klaene}, {Mahadevan}, {Mathur}, {Mosser}, {Muna}, {Munn}, {Nichol}, {O'Connell}, {Parejko}, {Robin}, {Rocha-Pinto}, {Schultheis}, {Serenelli}, {Shane}, {Silva Aguirre}, {Sobeck}, {Thompson}, {Troup}, {Weinberg}, \& {Zamora}}]{Majewski2017}
{Majewski}, S.~R., {Schiavon}, R.~P., {Frinchaboy}, P.~M., {et~al.} 2017, \aj, 154, 94

\bibitem[{{Malhan} {et~al.}(2022){Malhan}, {Ibata}, {Sharma}, {Famaey}, {Bellazzini}, {Carlberg}, {D'Souza}, {Yuan}, {Martin}, \& {Thomas}}]{Malhan22}
{Malhan}, K., {Ibata}, R.~A., {Sharma}, S., {et~al.} 2022, \apj, 926, 107

\bibitem[{{Massari} {et~al.}(2024){Massari}, {Dalessandro}, {Erkal}, {Balbinot}, {Bovy}, {McDonald}, {Ferguson}, {Larsen}, {Lan{\c{c}}on}, {Annibali}, {Goldman}, {Kuzma}, {Voggel}, {Saifollahi}, {Cuillandre}, {Schirmer}, {Kluge}, {Altieri}, {Amara}, {Andreon}, {Auricchio}, {Baldi}, {Balestra}, {Bardelli}, {Basset}, {Bender}, {Bonino}, {Branchini}, {Brescia}, {Brinchmann}, {Camera}, {Candini}, {Capobianco}, {Carbone}, {Carlberg}, {Carretero}, {Casas}, {Castellano}, {Cavuoti}, {Cimatti}, {Congedo}, {Conselice}, {Conversi}, {Copin}, {Corcione}, {Courbin}, {Courtois}, {Degaudenzi}, {Dinis}, {Dubath}, {Dupac}, {Dusini}, {Fabricius}, {Farina}, {Farrens}, {Ferriol}, {Frailis}, {Franceschi}, {Garilli}, {Gillis}, {Giocoli}, {Grazian}, {Guzzo}, {Hoar}, {Hoekstra}, {Holliman}, {Holmes}, {Hook}, {Hormuth}, {Hornstrup}, {Hudelot}, {Jahnke}, {Keih{\"a}nen}, {Kermiche}, {Kiessling}, {Kitching}, {Kohley}, {Kubik}, {K{\"u}mmel}, {Kunz}, {Kurki-Suonio}, {Ligori}, {Lilje}, {Lindholm}, {Lloro}, {Maino}, {Maiorano}, {Mansutti},
  {Marggraf}, {Markovic}, {Martinet}, {Marulli}, {Massey}, {Maurogordato}, {Medinaceli}, {Mei}, {Mellier}, {Meneghetti}, {Meylan}, {Moresco}, {Moscardini}, {Munari}, {Nakajima}, {Nichol}, {Niemi}, {Padilla}, {Paltani}, {Pasian}, {Pedersen}, {Percival}, {Pettorino}, {Pires}, {Polenta}, {Poncet}, {Popa}, {Pozzetti}, {Racca}, {Raison}, {Rebolo}, {Renzi}, {Rhodes}, {Riccio}, {Rix}, {Romelli}, {Roncarelli}, {Rossetti}, {Saglia}, {Sapone}, {Sartoris}, {Schneider}, {Schrabback}, {Secroun}, {Seidel}, {Seiffert}, {Serrano}, {Sirignano}, {Sirri}, {Skottfelt}, {Stanco}, {Tallada-Cresp{\'\i}}, {Teplitz}, {Tereno}, {Toledo-Moreo}, {Torradeflot}, {Tutusaus}, {Valenziano}, {Vassallo}, {Veropalumbo}, {Wang}, {Weller}, {Zacchei}, {Zamorani}, {Zoubian}, {Zucca}, {Bolzonella}, {Burigana}, {Morris}, {Scottez}, {Simon}, {Mart{\'\i}n-Fleitas}, \& {Scott}}]{Massari24}
{Massari}, D., {Dalessandro}, E., {Erkal}, D., {et~al.} 2024, arXiv e-prints, arXiv:2405.13498

\bibitem[{{Massari} {et~al.}(2019){Massari}, {Koppelman}, \& {Helmi}}]{Massari19}
{Massari}, D., {Koppelman}, H.~H., \& {Helmi}, A. 2019, \aap, 630, L4

\bibitem[{{McMillan}(2017)}]{McMillan17}
{McMillan}, P.~J. 2017, \mnras, 465, 76

\bibitem[{{Meng} \& {Gnedin}(2022)}]{Meng22}
{Meng}, X. \& {Gnedin}, O.~Y. 2022, \mnras, 515, 1065

\bibitem[{{Miyamoto} \& {Nagai}(1975)}]{1975PASJ...27..533M}
{Miyamoto}, M. \& {Nagai}, R. 1975, \pasj, 27, 533

\bibitem[{{Mori} {et~al.}(2024){Mori}, {Di Matteo}, {Salvadori}, {Khoperskov}, {Pagnini}, \& {Haywood}}]{2024A&A...690A.136M}
{Mori}, A., {Di Matteo}, P., {Salvadori}, S., {et~al.} 2024, \aap, 690, A136

\bibitem[{{Myeong} {et~al.}(2018){Myeong}, {Evans}, {Belokurov}, {Sanders}, \& {Koposov}}]{Myeong18}
{Myeong}, G.~C., {Evans}, N.~W., {Belokurov}, V., {Sanders}, J.~L., \& {Koposov}, S.~E. 2018, \apjl, 863, L28

\bibitem[{{Navarro} {et~al.}(1997){Navarro}, {Frenk}, \& {White}}]{Navarro97}
{Navarro}, J.~F., {Frenk}, C.~S., \& {White}, S. D.~M. 1997, \apj, 490, 493

\bibitem[{{Nelson} {et~al.}(2019{\natexlab{a}}){Nelson}, {Pillepich}, {Springel}, {Pakmor}, {Weinberger}, {Genel}, {Torrey}, {Vogelsberger}, {Marinacci}, \& {Hernquist}}]{Nelson19b}
{Nelson}, D., {Pillepich}, A., {Springel}, V., {et~al.} 2019{\natexlab{a}}, \mnras, 490, 3234

\bibitem[{{Nelson} {et~al.}(2019{\natexlab{b}}){Nelson}, {Springel}, {Pillepich}, {Rodriguez-Gomez}, {Torrey}, {Genel}, {Vogelsberger}, {Pakmor}, {Marinacci}, {Weinberger}, {Kelley}, {Lovell}, {Diemer}, \& {Hernquist}}]{Nelson19a}
{Nelson}, D., {Springel}, V., {Pillepich}, A., {et~al.} 2019{\natexlab{b}}, Computational Astrophysics and Cosmology, 6, 2

\bibitem[{{Newton} {et~al.}(2024){Newton}, {Davies}, {Pfeffer}, {Crain}, {Kruijssen}, {Pontzen}, \& {Bastian}}]{Newton24}
{Newton}, O., {Davies}, J.~J., {Pfeffer}, J., {et~al.} 2024, arXiv e-prints, arXiv:2409.04516

\bibitem[{{Pagnini} {et~al.}(2025{\natexlab{a}}){Pagnini}, {Di Matteo}, {Haywood}, {Mastrobuono-Battisti}, {Renaud}, {Mondelin}, {Agertz}, {Bianchini}, {Casamiquela}, {Khoperskov}, \& {Ryde}}]{Pagnini25}
{Pagnini}, G., {Di Matteo}, P., {Haywood}, M., {et~al.} 2025{\natexlab{a}}, \aap, 693, A155

\bibitem[{{Pagnini} {et~al.}(2025{\natexlab{b}}){Pagnini}, {Di Matteo}, {Haywood}, {Mastrobuono-Battisti}, {Renaud}, {Mondelin}, {Agertz}, {Bianchini}, {Casamiquela}, {Khoperskov}, \& {Ryde}}]{2025A&A...693A.155P}
{Pagnini}, G., {Di Matteo}, P., {Haywood}, M., {et~al.} 2025{\natexlab{b}}, \aap, 693, A155

\bibitem[{{Pagnini} {et~al.}(2023){Pagnini}, {Di Matteo}, {Khoperskov}, {Mastrobuono-Battisti}, {Haywood}, {Renaud}, \& {Combes}}]{Pagnini23}
{Pagnini}, G., {Di Matteo}, P., {Khoperskov}, S., {et~al.} 2023, \aap, 673, A86

\bibitem[{{Park} {et~al.}(2022){Park}, {Shin}, {Smith}, \& {Chun}}]{Park22}
{Park}, S.-M., {Shin}, J., {Smith}, R., \& {Chun}, K. 2022, \apj, 941, 91

\bibitem[{{Petts} {et~al.}(2016){Petts}, {Read}, \& {Gualandris}}]{2016MNRAS.463..858P}
{Petts}, J.~A., {Read}, J.~I., \& {Gualandris}, A. 2016, \mnras, 463, 858

\bibitem[{{Pfeffer} {et~al.}(2018){Pfeffer}, {Kruijssen}, {Crain}, \& {Bastian}}]{Pfeffer18}
{Pfeffer}, J., {Kruijssen}, J.~M.~D., {Crain}, R.~A., \& {Bastian}, N. 2018, \mnras, 475, 4309

\bibitem[{{Pfeffer} {et~al.}(2021){Pfeffer}, {Lardo}, {Bastian}, {Saracino}, \& {Kamann}}]{Pfeffer21}
{Pfeffer}, J., {Lardo}, C., {Bastian}, N., {Saracino}, S., \& {Kamann}, S. 2021, \mnras, 500, 2514

\bibitem[{{Pfeffer} {et~al.}(2020){Pfeffer}, {Trujillo-Gomez}, {Kruijssen}, {Crain}, {Hughes}, {Reina-Campos}, \& {Bastian}}]{Pfeffer20}
{Pfeffer}, J.~L., {Trujillo-Gomez}, S., {Kruijssen}, J.~M.~D., {et~al.} 2020, \mnras, 499, 4863

\bibitem[{{Pillepich} {et~al.}(2019){Pillepich}, {Nelson}, {Springel}, {Pakmor}, {Torrey}, {Weinberger}, {Vogelsberger}, {Marinacci}, {Genel}, {van der Wel}, \& {Hernquist}}]{Pillepich19}
{Pillepich}, A., {Nelson}, D., {Springel}, V., {et~al.} 2019, \mnras, 490, 3196

\bibitem[{{Pillepich} {et~al.}(2024){Pillepich}, {Sotillo-Ramos}, {Ramesh}, {Nelson}, {Engler}, {Rodriguez-Gomez}, {Fournier}, {Donnari}, {Springel}, \& {Hernquist}}]{Pillepich24}
{Pillepich}, A., {Sotillo-Ramos}, D., {Ramesh}, R., {et~al.} 2024, \mnras, 535, 1721

\bibitem[{{Pillepich} {et~al.}(2018){Pillepich}, {Springel}, {Nelson}, {Genel}, {Naiman}, {Pakmor}, {Hernquist}, {Torrey}, {Vogelsberger}, {Weinberger}, \& {Marinacci}}]{2018MNRAS.473.4077P}
{Pillepich}, A., {Springel}, V., {Nelson}, D., {et~al.} 2018, \mnras, 473, 4077

\bibitem[{{Ramos-Almendares} {et~al.}(2020){Ramos-Almendares}, {Sales}, {Abadi}, {Doppel}, {Muriel}, \& {Peng}}]{Ramos-Almendares20}
{Ramos-Almendares}, F., {Sales}, L.~V., {Abadi}, M.~G., {et~al.} 2020, \mnras, 493, 5357

\bibitem[{{Reina-Campos} {et~al.}(2022){Reina-Campos}, {Keller}, {Kruijssen}, {Gensior}, {Trujillo-Gomez}, {Jeffreson}, {Pfeffer}, \& {Sills}}]{ReinaCampos22}
{Reina-Campos}, M., {Keller}, B.~W., {Kruijssen}, J.~M.~D., {et~al.} 2022, \mnras, 517, 3144

\bibitem[{{Renaud}(2020{\natexlab{a}})}]{Renaud20}
{Renaud}, F. 2020{\natexlab{a}}, in IAU Symposium, Vol. 351, Star Clusters: From the Milky Way to the Early Universe, ed. A.~{Bragaglia}, M.~{Davies}, A.~{Sills}, \& E.~{Vesperini}, 40--46

\bibitem[{{Renaud}(2020{\natexlab{b}})}]{2020IAUS..351...40R}
{Renaud}, F. 2020{\natexlab{b}}, in IAU Symposium, Vol. 351, Star Clusters: From the Milky Way to the Early Universe, ed. A.~{Bragaglia}, M.~{Davies}, A.~{Sills}, \& E.~{Vesperini}, 40--46

\bibitem[{{Renaud} {et~al.}(2017{\natexlab{a}}){Renaud}, {Agertz}, \& {Gieles}}]{Renaud17}
{Renaud}, F., {Agertz}, O., \& {Gieles}, M. 2017{\natexlab{a}}, \mnras, 465, 3622

\bibitem[{{Renaud} {et~al.}(2017{\natexlab{b}}){Renaud}, {Agertz}, \& {Gieles}}]{2017MNRAS.465.3622R}
{Renaud}, F., {Agertz}, O., \& {Gieles}, M. 2017{\natexlab{b}}, \mnras, 465, 3622

\bibitem[{{Rodriguez} {et~al.}(2023){Rodriguez}, {Hafen}, {Grudi{\'c}}, {Lamberts}, {Sharma}, {Faucher-Gigu{\`e}re}, \& {Wetzel}}]{Rodriguez23}
{Rodriguez}, C.~L., {Hafen}, Z., {Grudi{\'c}}, M.~Y., {et~al.} 2023, \mnras, 521, 124

\bibitem[{{Saifollahi} {et~al.}(2025{\natexlab{a}}){Saifollahi}, {Lan{\c{c}}on}, {Cantiello}, {Cuillandre}, {Bethermin}, {Carollo}, {Duc}, {Ferr{\'e}-Mateu}, {Hatch}, {Hilker}, {Hunt}, {Marleau}, {Rom{\'a}n}, {S{\'a}nchez-Janssen}, {Tortora}, {Urbano}, {Voggel}, {Bolzonella}, {Bouy}, {Kluge}, {Schirmer}, {Stone}, {Giocoli}, {Knapen}, {Le}, {Mondelin}, {Aghanim}, {Altieri}, {Andreon}, {Auricchio}, {Baccigalupi}, {Bagot}, {Baldi}, {Balestra}, {Bardelli}, {Basset}, {Battaglia}, {Biviano}, {Bonchi}, {Bonino}, {Bon}, {Branchini}, {Brescia}, {Brinchmann}, {Camera}, {Capobianco}, {Carbone}, {Carretero}, {Casas}, {Castellano}, {Castignani}, {Cavuoti}, {Chambers}, {Cimatti}, {Colodro-Conde}, {Congedo}, {Conselice}, {Conversi}, {Copin}, {Courbin}, {Courtois}, {Cropper}, {Da Silva}, {Degaudenzi}, {De Lucia}, {Dole}, {Douspis}, {Dubath}, {Duncan}, {Dupac}, {Dusini}, {Escoffier}, {Farina}, {Farinelli}, {Faustini}, {Ferriol}, {Fotopoulou}, {Frailis}, {Franceschi}, {Fumana}, {Galeotta}, {George}, {Gillis}, {Gracia-Carpio},
  {Grazian}, {Grupp}, {Haugan}, {Hoar}, {Hoekstra}, {Holmes}, {Hook}, {Hormuth}, {Hornstrup}, {Jahnke}, {Jhabvala}, {Keih{\"a}nen}, {Kermiche}, {Kiessling}, {Kubik}, {K{\"u}mmel}, {Kunz}, {Kurki-Suonio}, {Lahav}, {Laureijs}, {Le Brun}, {Le Mignant}, {Ligori}, {Lilje}, {Lindholm}, {Lloro}, {Maino}, {Maiorano}, {Mansutti}, {Marggraf}, {Martinelli}, {Martinet}, {Marulli}, {Massey}, {Maurogordato}, {Medinaceli}, {Mei}, {Mellier}, {Meneghetti}, {Merlin}, {Meylan}, {Mora}, {Moresco}, {Moscardini}, {Nakajima}, {Neissner}, {Niemi}, {Padilla}, {Paltani}, {Pasian}, {Pedersen}, {Percival}, {Pettorino}, {Pires}, {Polenta}, {Poncet}, {Popa}, {Pozzetti}, {Raison}, {Rebolo}, {Renzi}, {Rhodes}, {Riccio}, {Romelli}, {Roncarelli}, {Saglia}, {Sakr}, {S{\'a}nchez}, {Sapone}, {Sartoris}, {Schewtschenko}, {Schneider}, {Schrabback}, {Seidel}, {Seiffert}, {Serrano}, {Sirignano}, {Sirri}, {Stanco}, {Steinwagner}, {Tallada-Cresp{\'\i}}, {Taylor}, {Tereno}, {Toft}, {Toledo-Moreo}, {Torradeflot}, {Tsyganov}, {Tutusaus}, {Valentijn},
  {Valenziano}, {Valiviita}, {Vassallo}, {Verdoes Kleijn}, {Veropalumbo}, {Wang}, {Weller}, {Zamorani}, {Zerbi}, {Zucca}, {Burigana}, {Mart{\'\i}n-Fleitas}, \& {Scottez}}]{Sai2025V1}
{Saifollahi}, T., {Lan{\c{c}}on}, A., {Cantiello}, M., {et~al.} 2025{\natexlab{a}}, arXiv e-prints, arXiv:2503.16367

\bibitem[{{Saifollahi} {et~al.}(2025{\natexlab{b}}){Saifollahi}, {Voggel}, {Lan{\c{c}}on}, {Cantiello}, {Raj}, {Cuillandre}, {Larsen}, {Marleau}, {Venhola}, {Schirmer}, {Carollo}, {Duc}, {Ferguson}, {Hunt}, {K{\"u}mmel}, {Laureijs}, {Marchal}, {Nucita}, {Peletier}, {Poulain}, {Rejkuba}, {S{\'a}nchez-Janssen}, {Urbano}, {Abdurro'uf}, {Altieri}, {Baes}, {Bolzonella}, {Conselice}, {Cote}, {Dimauro}, {Gonzalez}, {Habas}, {Hudelot}, {Kluge}, {Lonare}, {Massari}, {Romelli}, {Scaramella}, {Sola}, {Stone}, {Tortora}, {van Mierlo}, {Knapen}, {Mart{\'\i}n-Fleitas}, {Mora}, {Rom{\'a}n}, {Aghanim}, {Amara}, {Andreon}, {Auricchio}, {Baldi}, {Balestra}, {Bardelli}, {Basset}, {Bender}, {Bonino}, {Branchini}, {Brescia}, {Brinchmann}, {Camera}, {Capobianco}, {Carbone}, {Carretero}, {Casas}, {Castellano}, {Cavuoti}, {Cimatti}, {Congedo}, {Conversi}, {Copin}, {Courbin}, {Courtois}, {Cropper}, {Da Silva}, {Degaudenzi}, {Di Giorgio}, {Dinis}, {Dubath}, {Dupac}, {Dusini}, {Fabricius}, {Farina}, {Farrens}, {Ferriol}, {Fosalba},
  {Frailis}, {Franceschi}, {Fumana}, {Galeotta}, {Garilli}, {Gillard}, {Gillis}, {Giocoli}, {G{\'o}mez-Alvarez}, {Granett}, {Grazian}, {Grupp}, {Guzzo}, {Haugan}, {Hoar}, {Hoekstra}, {Holmes}, {Hook}, {Hormuth}, {Hornstrup}, {Jahnke}, {Jhabvala}, {Keih{\"a}nen}, {Kermiche}, {Kiessling}, {Kitching}, {Kohley}, {Kubik}, {Kuijken}, {Kunz}, {Kurki-Suonio}, {Lahav}, {Le Mignant}, {Ligori}, {Lilje}, {Lindholm}, {Lloro}, {Maino}, {Maiorano}, {Mansutti}, {Marggraf}, {Markovic}, {Martinet}, {Marulli}, {Massey}, {Maurogordato}, {McCracken}, {Medinaceli}, {Mei}, {Melchior}, {Mellier}, {Meneghetti}, {Meylan}, {Moresco}, {Moscardini}, {Munari}, {Nakajima}, {Nichol}, {Niemi}, {Padilla}, {Paltani}, {Pasian}, {Pedersen}, {Percival}, {Pettorino}, {Pires}, {Polenta}, {Poncet}, {Popa}, {Pozzetti}, {Racca}, {Raison}, {Rebolo}, {Refregier}, {Renzi}, {Rhodes}, {Riccio}, {Roncarelli}, {Rossetti}, {Saglia}, {Sapone}, {Sartoris}, {Schneider}, {Schrabback}, {Secroun}, {Seidel}, {Serrano}, {Sirignano}, {Sirri}, {Stanco},
  {Tallada-Cresp{\'\i}}, {Taylor}, {Teplitz}, {Tereno}, {Toledo-Moreo}, {Torradeflot}, {Tsyganov}, {Tutusaus}, {Valentijn}, {Valenziano}, {Vassallo}, {Verdoes Kleijn}, {Veropalumbo}, {Wang}, {Weller}, {Williams}, {Zamorani}, {Zucca}, {Biviano}, {Burigana}, {Scottez}, {Simon}, {Balogh}, \& {Scott}}]{Sai2025V2}
{Saifollahi}, T., {Voggel}, K., {Lan{\c{c}}on}, A., {et~al.} 2025{\natexlab{b}}, \aap, 697, A10

\bibitem[{{Sameie} {et~al.}(2023){Sameie}, {Boylan-Kolchin}, {Hopkins}, {Wetzel}, {Ma}, {Bullock}, {El-Badry}, {Quataert}, {Samuel}, {Schauer}, \& {Weisz}}]{Sameie23}
{Sameie}, O., {Boylan-Kolchin}, M., {Hopkins}, P.~F., {et~al.} 2023, \mnras, 522, 1800

\bibitem[{{Snaith} {et~al.}(2015){Snaith}, {Haywood}, {Di Matteo}, {Lehnert}, {Combes}, {Katz}, \& {G{\'o}mez}}]{Snaith2015}
{Snaith}, O., {Haywood}, M., {Di Matteo}, P., {et~al.} 2015, \aap, 578, A87

\bibitem[{{Sotillo-Ramos} {et~al.}(2022){Sotillo-Ramos}, {Pillepich}, {Donnari}, {Nelson}, {Eisert}, {Rodriguez-Gomez}, {Joshi}, {Vogelsberger}, \& {Hernquist}}]{Sotillo22}
{Sotillo-Ramos}, D., {Pillepich}, A., {Donnari}, M., {et~al.} 2022, \mnras, 516, 5404

\bibitem[{{Sun} {et~al.}(2023){Sun}, {Wang}, {Liu}, {Long}, {Chen}, \& {Gao}}]{Sun23}
{Sun}, G., {Wang}, Y., {Liu}, C., {et~al.} 2023, Research in Astronomy and Astrophysics, 23, 015013

\bibitem[{{Taylor} {et~al.}(2025){Taylor}, {Read}, {Orkney}, {Kim}, {Pontzen}, {Agertz}, {Rey}, {Andersson}, {Collins}, \& {Yates}}]{Taylor25}
{Taylor}, E.~D., {Read}, J.~I., {Orkney}, M. D.~A., {et~al.} 2025, arXiv e-prints, arXiv:2509.09582

\bibitem[{{Tsuchiya} {et~al.}(2003){Tsuchiya}, {Dinescu}, \& {Korchagin}}]{2003ApJ...589L..29T}
{Tsuchiya}, T., {Dinescu}, D.~I., \& {Korchagin}, V.~I. 2003, \apjl, 589, L29

\bibitem[{{Tsuchiya} {et~al.}(2004){Tsuchiya}, {Korchagin}, \& {Dinescu}}]{2004MNRAS.350.1141T}
{Tsuchiya}, T., {Korchagin}, V.~I., \& {Dinescu}, D.~I. 2004, \mnras, 350, 1141

\bibitem[{{van den Bergh}(2012)}]{vandenBergh2012}
{van den Bergh}, S. 2012, \apj, 746, 189

\bibitem[{{Vasiliev}(2019)}]{AGAMA}
{Vasiliev}, E. 2019, \mnras, 482, 1525

\bibitem[{{Vasiliev} \& {Baumgardt}(2021)}]{Vasiliev21}
{Vasiliev}, E. \& {Baumgardt}, H. 2021, \mnras, 505, 5978

\bibitem[{{Voggel} {et~al.}(2025){Voggel}, {Lan{\c{c}}on}, {Saifollahi}, {Larsen}, {Cantiello}, {Rejkuba}, {Cuillandre}, {Hudelot}, {Nucita}, {Urbano}, {Romelli}, {Raj}, {Schirmer}, {Tortora}, {Abdurro'uf}, {Annibali}, {Baes}, {Boldrini}, {Cabanac}, {Carollo}, {Conselice}, {Duc}, {Ferguson}, {Hunt}, {Knapen}, {Lonare}, {Marleau}, {Paolillo}, {Poulain}, {S{\'a}nchez-Janssen}, {Sola}, {Andreon}, {Auricchio}, {Baccigalupi}, {Baldi}, {Bardelli}, {Bodendorf}, {Bonino}, {Branchini}, {Brescia}, {Brinchmann}, {Camera}, {Capobianco}, {Carbone}, {Carlberg}, {Carretero}, {Casas}, {Castellano}, {Castignani}, {Cavuoti}, {Cimatti}, {Colodro-Conde}, {Congedo}, {Conversi}, {Copin}, {Courbin}, {Courtois}, {Cropper}, {Da Silva}, {Degaudenzi}, {De Lucia}, {Di Giorgio}, {Dinis}, {Dubath}, {Dupac}, {Dusini}, {Farina}, {Farrens}, {Ferriol}, {Fotopoulou}, {Frailis}, {Franceschi}, {Fumana}, {Galeotta}, {George}, {Gillard}, {Gillis}, {Giocoli}, {G{\'o}mez-Alvarez}, {Grazian}, {Grupp}, {Haugan}, {Hoekstra}, {Holmes}, {Hook},
  {Hormuth}, {Hornstrup}, {Jahnke}, {Keih{\"a}nen}, {Kermiche}, {Kiessling}, {Kilbinger}, {Kohley}, {Kubik}, {K{\"u}mmel}, {Kunz}, {Kurki-Suonio}, {Laureijs}, {Liebing}, {Ligori}, {Lilje}, {Lindholm}, {Lloro}, {Maino}, {Maiorano}, {Mansutti}, {Marggraf}, {Markovic}, {Martinelli}, {Martinet}, {Marulli}, {Massey}, {Maurogordato}, {Medinaceli}, {Mei}, {Mellier}, {Meneghetti}, {Merlin}, {Meylan}, {Moresco}, {Moscardini}, {Munari}, {Nakajima}, {Neissner}, {Nichol}, {Niemi}, {Nightingale}, {Padilla}, {Paltani}, {Pasian}, {Pedersen}, {Pettorino}, {Pires}, {Polenta}, {Poncet}, {Popa}, {Pozzetti}, {Raison}, {Rebolo}, {Renzi}, {Rhodes}, {Riccio}, {Roncarelli}, {Rossetti}, {Saglia}, {Sakr}, {Sapone}, {Sartoris}, {Scaramella}, {Schneider}, {Schrabback}, {Secroun}, {Sefusatti}, {Seidel}, {Serrano}, {Sirignano}, {Sirri}, {Stanco}, {Steinwagner}, {Surace}, {Tallada-Cresp{\'\i}}, {Teplitz}, {Tereno}, {Toledo-Moreo}, {Torradeflot}, {Tutusaus}, {Valentijn}, {Valenziano}, {Vassallo}, {Veropalumbo}, {Wang}, {Weller}, {Zamorani},
  {Zucca}, {Biviano}, {Bolzonella}, {Bozzo}, {Burigana}, {Calabrese}, {Di Ferdinando}, {Escartin Vigo}, {Farinelli}, {Gracia-Carpio}, {Mauri}, {Scottez}, {Tenti}, {Viel}, {Wiesmann}, {Akrami}, {Allevato}, {Anselmi}, {Ballardini}, {Bethermin}, {Blanchard}, {Blot}, {Borgani}, {Borlaff}, {Bruton}, \& {Calabro}}]{Voggel25}
{Voggel}, K., {Lan{\c{c}}on}, A., {Saifollahi}, T., {et~al.} 2025, \aap, 693, A251

\bibitem[{{Weinberger} {et~al.}(2017){Weinberger}, {Springel}, {Hernquist}, {Pillepich}, {Marinacci}, {Pakmor}, {Nelson}, {Genel}, {Vogelsberger}, {Naiman}, \& {Torrey}}]{2017MNRAS.465.3291W}
{Weinberger}, R., {Springel}, V., {Hernquist}, L., {et~al.} 2017, \mnras, 465, 3291

\end{thebibliography}

\appendix

\section{Impact of free parameters of our model}

\begin{table*}[!]
\begin{center}
\label{tab:landscape}
\caption{Impact of free parameters in our modelling}
\begin{tabular}{cccccccccccc}
 \hline
  & Ex-situ in 0.95 boundary & In-situ in 0.05 boundary & $N_{\rm ex-situ}$ & $N_{\rm mixing\; zone}$ & $N_{\rm in-situ}$ \\
  & $\%$ & $\%$ & \\
  \hline
  Fiducial model & 81.21 & 76.4 & 67 & 38 & 55 & \\
  Fixed disk potential & 81.28 & 74.97 & 67 & 40 & 53 & \\
  In-situ tagged at $z=3$ & 83.13 & 23.45 & 78 & 68 & 14\\
  $M_{\rm GC}^{\rm ini}  = 5\times 10^5$ $\rm{M_{\odot}}$ & 80.27 & 41 & 65 & 30 & 65 \\
  $M_{\rm GC}^{\rm ini} = 5\times 10^6$ $\rm{M_{\odot}}$ & 81.38 & 96.5 & 68 & 49 & 43  \\
  Mass loss \\
  ($\gamma = 0.7$ $\&$ $t_{0,\odot} = 10.7$ Myr) & 80.83 & 54.17 & 66 & 32 & 62 \\
    \hline
\end{tabular}
\label{tab1}
\parbox{\hsize}{Notes: From left to right, the columns report, for each configuration: the percentage of simulated GCs classified as ex-situ and in-situ using the 0.95 and 0.05 decision boundaries, respectively, as well as the number of GCs identified as ex-situ, in the mixing zone, and in-situ. Our fiducial model assumes $10^{6}~\rm{M_{\odot}}$ GCs tagged at $z=2$ for in-situ clusters and at $z=3$ for ex-situ ones, evolved in MW-like potentials without a disk component, with mass loss parametrized by $\gamma = 0.62$ and $t_{0,\odot} = 21.3$ Myr. The fixed disk is modeled using a \cite{1975PASJ...27..533M} potential with a mass of $6.8\times 10^{10}~\rm{M_{\odot}}$, a scale length $a = 3$ kpc, and a scale height $b = 280$ pc, consistent with the \texttt{MWPotential2014} used in \citet{Bovy15}.}
\end{center}
\end{table*}

\section{Orbital integrations of satellites}

\begin{figure}
\centering
\includegraphics[width=\hsize]{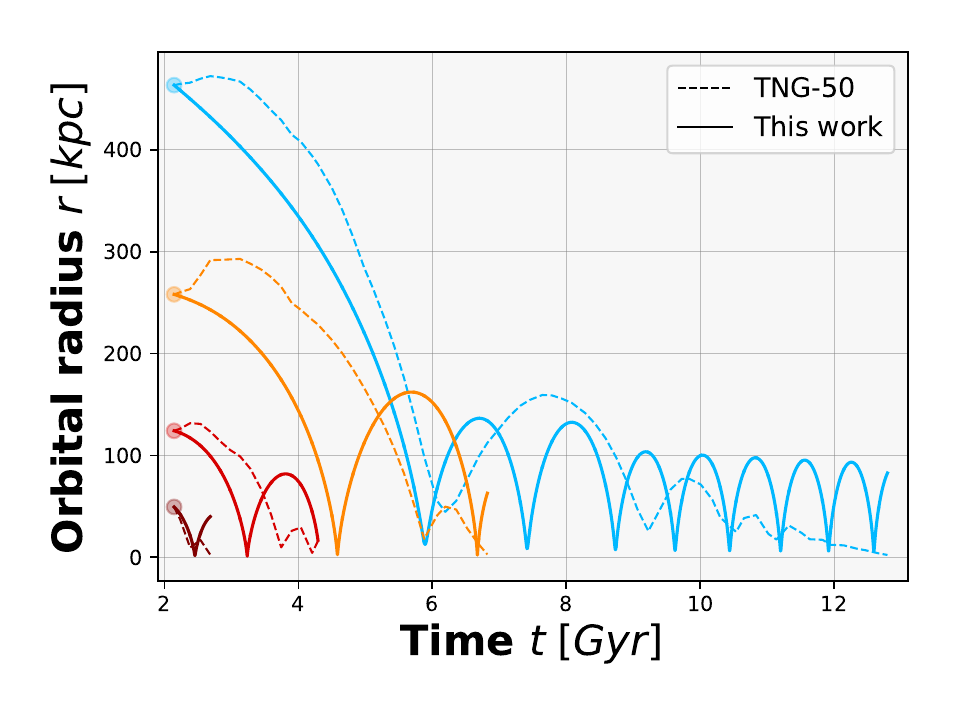}
\caption{Orbits of merging satellites from TNG50 and from our orbital integrations in our analytic MW potential. We have re-integrated satellite orbits both to improve the time resolution of orbits that are sometimes highly jagged in TNG50 due to non-linear time evolution, and to incorporate them as "moving potentials" in our full MW potential.}
\label{figA0}
\end{figure}

\end{document}